\documentclass[fleqn,usenatbib]{mnras}

\usepackage{comment}
\usepackage{rotating}
\usepackage{xcolor}
\usepackage{float}
\usepackage[T1]{fontenc}
\usepackage{placeins}

\DeclareRobustCommand{\VAN}[3]{#2}
\let\VANthebibliography\thebibliography
\def\thebibliography{\DeclareRobustCommand{\VAN}[3]{##3}\VANthebibliography}

\usepackage{amsmath}	
\usepackage{times}
\usepackage{verbatim}
\usepackage{graphicx,bm,amssymb}
\usepackage{epsfig}
\usepackage{amssymb}
\usepackage{natbib}
\usepackage{pstricks}
\usepackage{psfrag}
\usepackage[caption = false]{subfig}
\usepackage{newtxtext,newtxmath}
\usepackage{textcomp}

\title[GRB 191016A, max. 45 characters]{GRB 191016A: The onset of the forward shock and evidence of late energy injection}

\author[Pereyra et al.]{%
M. Pereyra$^{1}$\thanks{E-mail: mpereyra@astro.unam.mx},
N.~Fraija$^{2}$,
A.~M.~Watson$^{2}$,
R.~L.~Becerra$^{3}$,
N.~R.~Butler$^{4}$,
F.~De Colle$^{3}$, 
E.~Troja$^{5,6}$, \newauthor
E.~Fraire-Bonilla$^{7}$,
S.~Dichiara$^{5,6}$,
W.~H.~Lee$^{2}$,
A.~S.~Kutyrev$^{6}$, 
J.~X.~Prochaska$^{8}$,
J.~S.~Bloom$^{10}$, \newauthor
J.~J.~Gonz\'alez$^{2}$,
E.~Ramirez-Ruiz$^{8}$,
and M.~G.~Richer$^{9}$
\\
$^1$ CONACYT, Instituto de Astronomía, Universidad Nacional Autónoma de México, Unidad Acad\'emica en Ensenada, 22860 Ensenada, BC, Mexico\\
$^2$ Instituto de Astronom{\'\i}a, Universidad Nacional Aut\'onoma de M\'exico, Apartado Postal 70-264, 04510 M\'exico, CDMX, Mexico\\
$^3$ Instituto de Ciencias Nucleares, Universidad Nacional Aut\'onoma de M\'exico, Apartado Postal 70-264, 04510 M\'exico, CDMX, Mexico\\
$^4$ School of Earth and Space Exploration, Arizona State University, Tempe, AZ 85287, USA\\
$^5$ Department of Astronomy, University of Maryland, College Park, MD 20742-4111, USA\\
$^6$ Astrophysics Science Division, NASA Goddard Space Flight Center, 8800 Greenbelt Road, Greenbelt, MD 20771, USA\\
$^7$ Facultad de Ciencias F\'isico Matem\'aticas, Universidad Autónoma de Nuevo León, M\'exico.\\
$^8$ Department of Astronomy and Astrophysics, UCO/Lick Observatory, University of California, 1156 High Street, Santa Cruz, CA 95064, USA\\
$^9$ Instituto de Astronomía, Universidad Nacional Autónoma de México, Unidad Acad\'emica en Ensenada, 22860 Ensenada, BC, Mexico\\
$^{10}$ Department of Astronomy, University of California, Berkeley, CA 94720-3411, USA\\
}

\date{Accepted XXX. Received YYY; in original form ZZZ}

\pubyear{2015}

\begin{document}
\label{firstpage}
\pagerange{\pageref{firstpage}--\pageref{lastpage}}
\maketitle

\begin{abstract}
We present optical and near-infrared photometric observations of GRB 191016 with the COATLI, DDOTI and RATIR ground-based telescopes over the first three nights. We present the temporal evolution of the optical afterglow and describe 5 different stages that were not completely characterized in previous works, mainly due to scarcity of data points to accurately fit the different components of the optical emission. After the end of the prompt gamma-ray emission, we observed the afterglow {\itshape rise} slowly in the optical and near-infrared (NIR) wavelengths and peak at around $T+1450$~s in all filters. This was followed by an {\itshape early decay}, a clear {\itshape plateau} from $T+5000$~s to $T+11000$~s, and then a regular {\itshape late decay}. We also present evidence of the {\itshape jet break} at later times, with a temporal index in good agreement with the temporal slope obtained from X-ray observations. Although many of the features observed in the optical light curves of GRBs are usually well explained by a reverse shock (RS) or forward shock (FS), the shallowness of the optical rise and enhanced peak emission in the GRB191016A afterglow is not well-fitted by only a FS or a RS. We propose a theoretical model which considers both of these components and combines an evolving FS with a later embedded RS and a subsequent late energy injection from the central engine activity. We use this model to successfully explain the temporal evolution of the light curves and discuss its implications on the fireball properties.
\end{abstract}



\section{Introduction}

Gamma-ray bursts (GRBs) are highly energetic transient sources, emitting an isotropic energy of 10$^{50}$-10$^{53}$ ergs in $\gamma$-rays. They are associated with catastrophic events involving core-collapse supernovae and mergers of compact objects such as neutron stars (NSs) and black holes (BHs). 

The duration of a GRB is characterized by the time during which 50 or 90 percent of the total energy above the background level is detected, which is commonly referred as T$_{50}$  and T$_{90}$, respectively \citep{1993ApJ...413L.101K}. According to their duration, GRBs can be broadly classified as either long or short. Short GRBs, with a duration $\lesssim$ 2~s, are thought to be produced by the coalescence of binary neutron stars, while long GRBs, with durations $\gtrsim$ 2~s, are usually associated to the death of massive stars (see, e.g., \citealp{2006RPPh...69.2259M,2009grbb.book.....V, 2015PhR...561....1K} for reviews).

The multi-wavelength emission observed from both types of GRBs has been successfully modelled within the framework of the \emph{fireball} model. This model predicts the ejection of a relativistic jet, the fireball, during the formation of a black hole \citep{1992MNRAS.258P..41R,1997ApJ...476..232M,1998ApJ...503..314P,1998ApJ...497L..17S}.  
The prompt $\gamma$-ray emission is usually attributed to internal shocks produced when shells of material, thrown violently from the progenitor at different Lorentz factors, overtake each other. The high-energy prompt emission is followed by the afterglow, produced when the outflow is decelerated by the external medium and the shocked material radiates at X-ray and lower frequencies.

Since launch, the Neil Gehrels Swift Observatory satellite has observed a large sample of early X-ray and ultraviolet/optical light curves with its X-ray Telescope (XRT; \cite{2005SSRv..120..143B}) and Ultraviolet Optical Telescope (UVOT; \citealp{2005SSRv..120...95R}). Despite of the large number of GRBs detected to date \citep{2020ApJ...893...77W,2020ApJ...893...46V}, the early stages of the afterglow are not well understood. The early afterglow is thought to connect the tail of the prompt emission to the subsequent afterglow. Therefore, it becomes particularly useful to answer fundamental questions such as: What is the connection between the prompt emission and the afterglow? What is the density stratification of the circumburst medium? What is the initial Lorentz factor of the fireball? What is the role of the reverse shock? Does the central engine activity actually stop after the prompt emission is over? \citep{2006ApJ...642..354Z}.

Addressing these questions requires observing both the early and later stages of the GRB afterglow to better understand its global properties.
Rapid response telescopes and  satellites, whose capabilities allow observations within a few seconds of an alert, are the best instruments to observe the afterglow quickly, during the prompt gamma emission and/or just after its end. In most GRBs, the observed evolution of the X-ray afterglow may be summarized as follows: a steep decay phase related with the end of the prompt emission phase, a shallow decay phase or plateau, a “normal” decay phase, a jet-break steepening, and flares in some but not all bursts (see, e.g., \citealp{2006ApJ...642..354Z,2006ApJ...642..389N,2009MNRAS.397.1177E}). The optical light curve of the afterglow is often well described with a single shallow decay phase, but also particular features, such as bumps, plateaus, and late rebrightenings have been observed in some bursts (see, e.g., \citealp{2009MNRAS.395..490O,2010ApJ...725.2209L,2011MNRAS.414.3537P,2013A&A...557A..12Z,2017ApJ...844...79Y,2017ApJS..228...13R} and references therein). 

The long GRB 191016A was detected by Swift/BAT at 04:09:00 UTC on 2019 October 16 \citep{2019GCN.26008....1G}, and its long-lasting afterglow emission was extensively observed in optical wavelengths by our ground-based robotic telescopes RATIR, COATLI, and DDOTI 
\citep{2012SPIE.8446E..10B,2012SPIE.8444E..5LW,2016SPIE.9908E..5OW,2016SPIE.9910E..0GW,2016SPIE.9908E..5QC} at the Observatorio Astron\'omico Nacional on the Sierra de San Pedro M\'artir in Baja California in Mexico. In \cite{2019GCN.26010....1W}, we reported the discovery of the afterglow with COATLI and its rise and subsequent fade. \cite{2021ApJ...911...43S} recently reported that the GRB 191016A did not trigger {\itshape Fermi}/GBM, but was detected in a more sensitive targeted search. \cite{2021ApJ...911...43S} also report TESS photometric observations. Although the TESS light curve has good photometric precision of 0.1-1\% at the observed peak brightness of the GRB, its monitoring cadence ranging from 10 to 30 minutes does not allow for a complete characterization of the evolution of the optical afterglow. Additional optical photometry was very recently presented by \cite{2021arXiv211109123S}, also limited around the peak time, with interesting polarization measurements that favours energy injection into the blast wave at late times. Although \cite{2021arXiv211109123S} provided a complete photometric and polarimetric data sets over the time interval $3987-7587$~s after trigger time, the presence of a probable jet break at later times is not strongly supported by the few data used in their analysis.

In this work, we present our  follow-up optical and near-infrared observations of the afterglow emission of this long GRB, started only 6 min after trigger and continued over the first three nights, with a typical cadence in the range of 5 to 80 sec. In Section~\ref{sec:Observations} we describe the observations and data acquisition. In Section~\ref{sec:Results} we present the physical properties of the afterglow, derived from our photometry. In Section~\ref{sec:Discussion} we discuss the scenarios that could explain the optical light curve behaviour of the GRB 191016A afterglow emission and present the theoretical model that successfully reproduces our whole data set. We finally present our conclusions in Section~\ref{sec:Conclusions}.


\section{Observations} \label{sec:Observations}

\subsection{Neil Gehrels Swift Observatory}

The {\itshape Swift}/BAT instrument \citep{2005SSRv..120..143B} triggered on GRB 191016A at $T =$ 2019 October 16 04:09:00.9 UTC (trigger 929744) \citep{2019GCN.26008....1G}. \citet{2019GCN.26012....1B} reported a total duration $T_{90}= 220 \pm 183$~s, with emission present from $T-40$~s to $T+420$~s and peaks at $T-10$~s and $T+35$~s. The analysis linked to by these authors gives $T_{50} = 61 \pm 11$~s. This firmly places GRB 191016A in the distribution of long GRBs.

Due to a Moon observing constraint, the Swift/XRT and \emph{Swift}/UVOT instruments did not observe the field until 12.5~h after the BAT trigger. Therefore, initially, the best positions based on high-energy emission were the on-board BAT position \citep{2019GCN.26008....1G} with a 90\% uncertainty of 3~arcmin in radius followed by the ground BAT position of 02:01:04.4 +24:30:30.4 (J2000) with a 90\% uncertainty of 1.5 arcmin in radius \citep{2019GCN.26012....1B}.

\cite{2019GCN.26026....1P} reported observations with the \emph{Swift}/XRT instrument from $T+12.5$~h to $T+15.8$~h and the detection of an uncatalogued X-ray source at 02:01:04.64 +24:30:35.6 (J2000) with an uncertainty of 1.9 arcsec in radius, consistent with the Swift/BAT position and the previously reported ground-based afterglow detection \citep{2019GCN.26010....1W}. The X-ray light curve at these later times can be modelled with a power-law decay for a temporal index $\alpha = 1.9$.

\cite{2019GCN.26024....1S} reported observations with the \emph{Swift}/UVOT instrument from $T+12.5$~h to $T+15.8$~h and the preliminary detection of a source at about 21 mag, in the white filter, consistent with both the XRT position and the previously reported ground-based afterglow detection. 

\subsection{Fermi}

 \cite{2021ApJ...911...43S} recently reported the detection of GRB 191016A with {\itshape Fermi}/GBM. The burst did not trigger the instrument, but was detected in a more sensitive targeted search. Their 50-300 keV light curve shows emission over about 100 seconds, again with two peaks.
 
\subsection{COATLI}

COATLI is a 50-cm robotic telescope installed at the Observatorio Astronómico Nacional on the Sierra de San Pedro Mártir in Baja California, Mexico \citep{2016SPIE.9908E..5OW,2016SPIE.9908E..5QC}. COATLI has an ASTELCO 50-cm Richey-Crétien telescope on a fast ASTELCO NTM-500 German equatorial mount and currently operates with an interim instrument, a CCD with a field of view of $12.8 \times 8.7$ arcmin and \emph{BVRIw} filters. 

COATLI received the GCN Notice with the initial BAT position at 2019 October 16 04:14:57.7 and automatically slewed to observe. The first exposure started at 04:15:19.6 or $T+378.7$~s. The delay between the trigger and our first exposure is due to an unexplained delay of 296.8~s between the trigger and the reception of the GCN Notice and 21.9~s for the slew, longer than in some other observations because the German equatorial mount had to flip from west to east.

On the first night, 2019 October 16, COATLI observed the field from just after the alert at 04:15 UTC until the end of morning astronomical twilight at 12:36 UTC. On the second night, 2019 October 17, COATLI observed the field from shortly after the start of evening astronomical twilight at 02:25 UTC until 11:21 UTC. The exposures were in the $w$ filter for a total exposure of 6.64~h and 5.32~h, respectively.

The observations were reduced and analysed by our real-time pipeline, which performs bias subtraction and twilight flat division, then uses astrometry.net \citep{2010AJ....139.1782L} for image alignment, {\sc swarp} \citep{2010ascl.soft10068B}  for image coaddition, and estimates the sky iteratively. Photometry is obtained by running {\sc sextractor} \citep{1996A&AS..117..393B} with a range of aperture diameters. A weighted average of the flux in this set of apertures for all stars in a given field is then used to construct an annular point-spread function (PSF) whose core-to-halo ratio is allowed to vary smoothly over the field. This PSF is then fitted to the annular flux values for each source to optimize the signal-to-noise for point source photometry. The $w$ magnitudes were calibrated against the Pan-STARRS1 catalogue using the empirical transformation from $g$ and $r$ given by \cite{2019ApJ...872..118B}, are on an approximate AB system, and are not corrected for Galactic extinction.

\cite{2019GCN.26010....1W} reported the first detection of the afterglow of GRB 191016A, which in our initial 40~m of COATLI observations appeared as an uncatalogued source at 02:01:04.75 +24:30:36.8 (J2000), consistent with the BAT position, that brightened from $w \approx 16.5$ to $w \approx 15.0$ before fading. The nature of this source was confirmed by subsequent observations of the optical light curve reported by \cite{2019GCN.26011....1Z}, \cite{2019GCN.26015....1W}, \cite{2019GCN.26017....1H}, \cite{2019GCN.26018....1K}, \cite{2019GCN.26019....1T}, \cite{2019GCN.26024....1S}, \cite{2019GCN.26026....1P}, \cite{2019GCN.26028....1M}, and \cite{2019GCN.26176....1S} and, in particular, the \emph{Swift}/XRT detection reported by \cite{2019GCN.26026....1P}.
 
We performed variable binning of the COATLI data to balance sampling and signal-to-noise. The 5~s exposures up to $T+2000$ s are binned in groups of 6 consecutive exposures taken over about 54~s. The  30~s exposures from $T + 2000$~s to $T + 2500$~s are not binned. The 30~s exposures from $T + 2500$~s to $T + 3500$~s are binned in groups of 4 consecutive exposures taken over about 136~s. The 30~s exposures from $T + 3500$~s to the end of the first night are in groups of 16 consecutive exposures taken over about 544~s. All the 30~s exposures taken on the seconds night are binned into a single 19050~s exposure taken over about 5.32~h.

The COATLI photometry is presented in Figure~\ref{fig:OAN_Follow-up} and Appendix~\ref{Tables_photometry}.

\subsection{RATIR}

RATIR is a four-channel imager mounted on the robotic Harold L.\ Johnson 1.5-meter Telescope of the Observatorio Astronómico Nacional on the Sierra de San Pedro Mártir in Baja California, Mexico \citep{2012SPIE.8446E..10B,2012SPIE.8444E..5LW}. The two CCD channels observe in $gr$ and $i$ and the two H2RG channels observe in $ZY$ and $JH$.

RATIR received the GCN Notice with the initial BAT position at 2019 October 16 04:14:57.7 UTC and automatically slewed to observe. The first exposure started at 04:16:25.9 UTC or $T+445.0$~s. The delay was slightly longer than for COATLI, since the Johnson Telescope slews slowly.

On the first night, 2019 October 16, RATIR observed the field from just after the alert at 04:16 UTC until the end of morning astronomical twilight at 12:36 UTC, initially in the $riZYJH$ filters, but from 05:58 UTC the $g$ filter was alternated with $r$.  The exposures were 80~s in $gri$ and 67~s in $ZYJH$. The total exposure was 2.27~h in $g$, 3.38~h in $r$, 5.64~h in $i$ and only 0.54~h in each of $ZYJH$, due to failures in the cryostat cooling system. 
On the second night, 2019 October 17, RATIR observed the field with only $gri$ filters. The total exposure was 0.80~h in each of $gr$ and 1.60~h in $i$.
On the third night, 2019 October 18, RATIR observed the field from shortly after the start of evening astronomical twilight at 02:48 UTC until 12:38 UTC. Only $ri$ CCD channels were used, amounting to a total exposure of 6.98~h in each band.

The observations were reduced and analysed by our real-time pipeline, which is similar to the COATLI pipeline. The photometry in $griZ$ was calibrated against the SDSS DR9 \citep{2012ApJS..203...21A} and the photometry in $YJH$ against 2MASS \citep{2006AJ....131.1163S}, using $JH$ to estimate $Y$ \citep{2007A&A...467..777C,2009MNRAS.394..675H}. The final photometry, presented in Figure \ref{fig:OAN_Follow-up}, is in an AB system and is not corrected for Galactic extinction in the direction of the GRB. RATIR data obtained from observations performed on 2019 October 17 and 2019 October 18 were grouped into one single bin per filter used. We grouped RATIR-\emph{r} data in 2880~s and 11280~s bins, while RATIR-\emph{i} data were grouped into 5600~s and 11280~s bins, for each night respectively.

The RATIR photometry is presented in Figure~\ref{fig:OAN_Follow-up} and Appendix~\ref{Tables_photometry}.

\subsection{DDOTI}
 
DDOTI is a wide-field, optical, robotic imager located at the Observatorio Astronómico Nacional on the Sierra de San Pedro Mártir in Baja California, Mexico \citep{2016SPIE.9910E..0GW}. It has an ASTELCO Systems NTM-500 mount with six Celestron RASA 28-cm astrographs each with an unfiltered Finger Lakes Instrumentation ML50100 front-illuminated CCD detector, an adapter of our own design and manufacture that allows static tip-tilt adjustment of the detector, and a modified Starlight Instruments motorized focuser. Each telescope has a field of about $3.4 \times 3.4$~deg with 2.0 arcsec pixels. The individual fields are arranged on the sky in a $2 \times 3$ grid to give a total field of 69~$\mathrm{deg}^2$.

DDOTI is designed to for observations of poorly localized event detected by Fermi/GBM and LIGO/Virgo \citep{2021MNRAS.507.1401B}. Nevertheless, it also responds to Swift/BAT events as a backup for COATLI and RATIR. DDOTI received the GCN Notice with the initial BAT position at 2019 October 16 04:14:57.7 UTC and automatically slewed to observe. The first exposure started at 04:21:44.1 UTC or $T+763.2$~s. The delay was longer than for COATLI, since DDOTI currently needs to refocus after a slew.

On the first night, 2019 October 16, DDOTI observed the field from just after the alert at 04:21 UTC until the end of morning astronomical twilight at 12:36 UTC. The exposures were in the $w$ filter and were 30~s with a cadence of about 40~s until 04:34 UTC and then were of 60~s with a cadence of about 70~s. The total exposure was 5.33~h. 

\begin{figure*}
\centering
    \includegraphics[width=\textwidth]{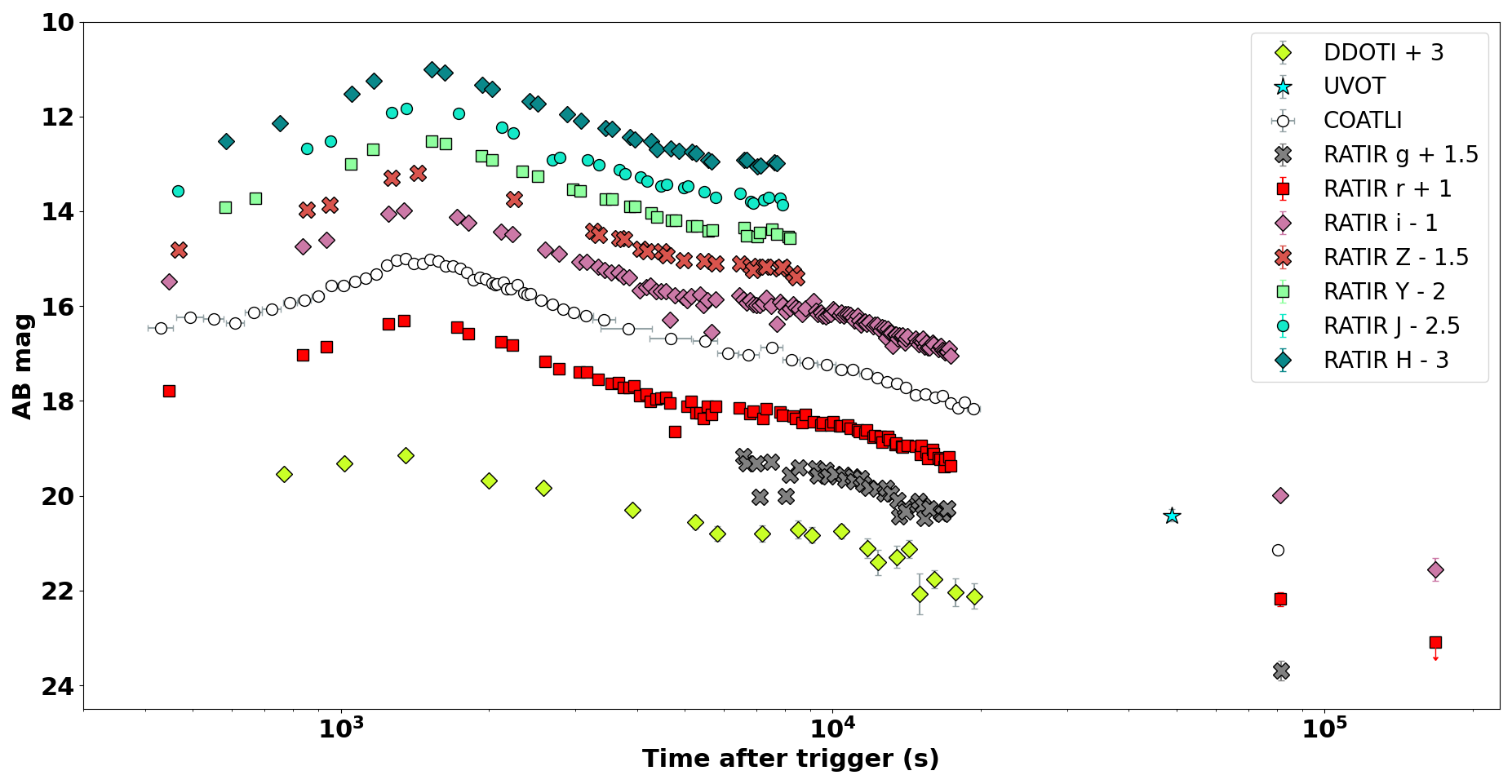}
    \caption{\textbf{Optical follow-up of the GRB 191016A afterglow emission.} Different symbols correspond to COATLI (white circles), RATIR (coloured squares, circles and diamonds) and DDOTI (yellow diamonds). DDOTI and RATIR data points were shifted, as described in the legend, only for illustrative purposes. Upper limits are indicated with arrows. We plot the UVOT observations with a star in cyan.}
    \label{fig:OAN_Follow-up}
\end{figure*}

The observations were reduced and analysed by our real-time pipeline, which is similar to the COATLI pipeline. The main differences are that the {\sc swarp} is used to estimate the sky background and the PSF is determined on a grid over the field of each CCD. The photometry in $w$ is calibrated against the APASS DR10 catalogue \citep{2018AAS...23222306H} using our measured transformation of $w \approx r + 0.23(g - r)$, is on an approximate AB system, and is not corrected for Galactic extinction in the direction of the GRB. Binned data from DDOTI photometry are plotted in Fig \ref{fig:OAN_Follow-up} as yellow diamonds. We group data from 4 consecutive 30~s exposures in 120~s bins up to $T + 1500$~s. For times between $T + 1500$~s and $T + 16000$~s, we combine data from 4 consecutive 60~s exposures in bins of 240~s. From $T + 16000$~s to the end of the first night we used time bins of 720~s, grouping 12 consecutive 60~s exposures.

The DDOTI photometry is presented in Figure~\ref{fig:OAN_Follow-up} and Appendix~\ref{Tables_photometry}.

\section{Results} \label{sec:Results}

In this section, we interpret the $\gamma$-ray observations of GRB 191016A in the prompt phase emission as well as the optical and NIR light curves during the afterglow phase. 

\subsection{Temporal analysis} 

\begin{figure*}
\centering
    \includegraphics[width=\textwidth]{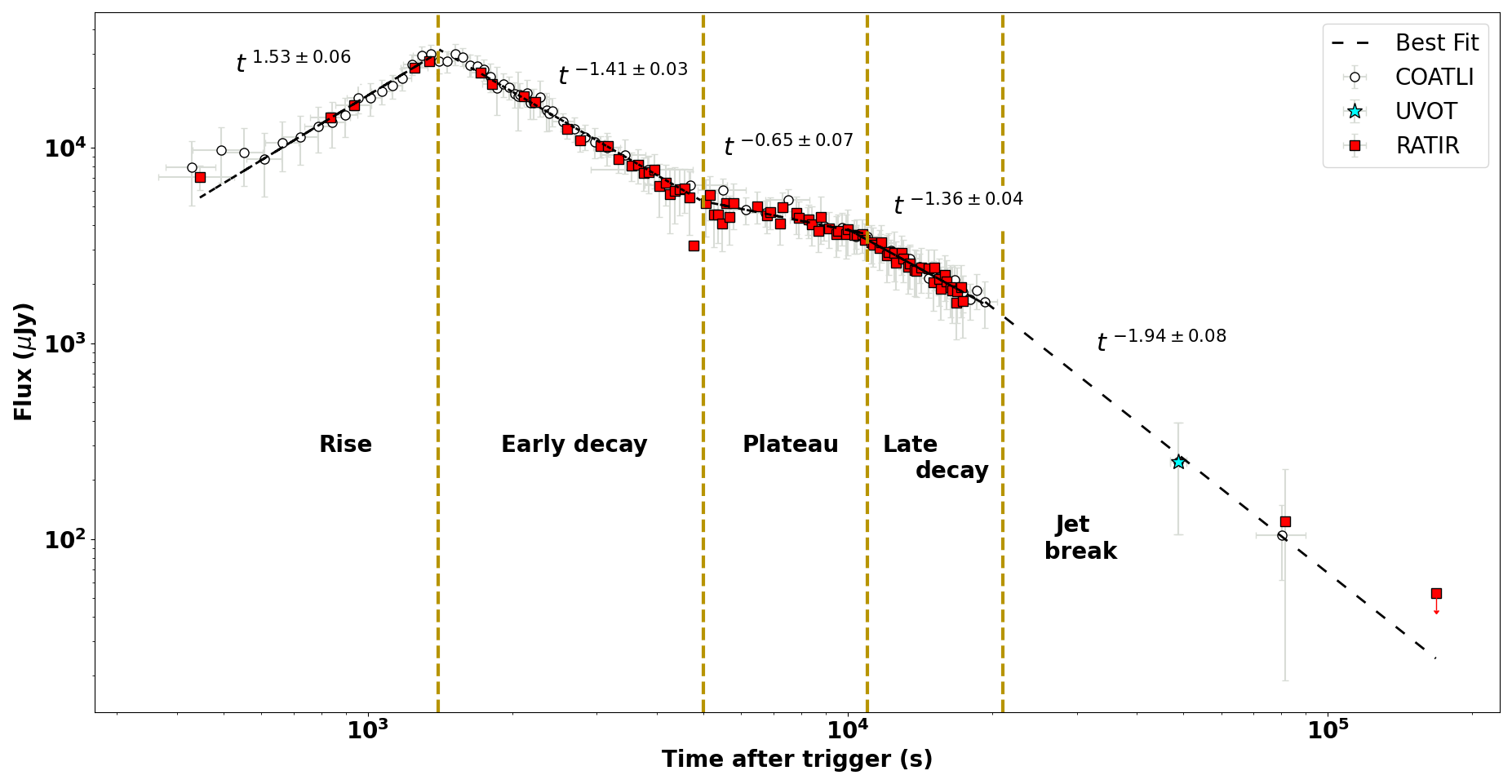}
    \caption{\textbf{Temporal evolution of GRB 191016A afterglow emission.} COATLI $w$ data are plotted using white circles, whereas RATIR $r$ data are shown in red squares. The star in cyan, at 12 h after trigger, is the UVOT \emph{w} measurement. The best fit for each power law segment is represented by the dashed line in black. Upper limits are indicated with arrows. Labels at the bottom of the plot indicate the different stages we identify along the afterglow evolution with the corresponding temporal index show above each power law segment fitted.}%
    \label{fig:PowerLaw_Fits}
\end{figure*}

To determine the temporal evolution of the optical afterglow we used COATLI $w$, UVOT $w$, and RATIR $r$ data, presented in Figure \ref{fig:PowerLaw_Fits}. We fitted the observed light curves with different power-law segments, assuming the following flux convention: $F \propto t^{\alpha}$, in which $F$ is the flux density, $t$ is the time since the BAT trigger, and $\alpha$ is the temporal index. 

Our observations show the optical emission rising from $T+600$~s. The flux increases with $\alpha_{O,\rm Rise}=1.53$, until it reaches a maximum at $t_{\rm peak}=1450$~s. An early decay in the light curve is observed afterwards, with a temporal index $\alpha_{O,\rm Early}=-1.41$. From $T+5000$~s to $T+11000$~s, there is a clear plateau with a shallower temporal index of $\alpha=-0.65$. After $T+11000$~s the flux continues decreasing in a regular decay with $\alpha_{O,\rm Late}=-1.36$. We also confirm the existence of a jet break observed at later times, with a temporal slope $\alpha_{O,\rm Break}=-1.94$ consistent with X-ray data, that occurs between 5.8 to 12.5 hrs after BAT trigger. The temporal index for each power-law segment of the afterglow emission are presented in Table \ref{table:indexes}.

\begin{table}
	\centering
	\caption{Temporal Power-Law Indices} 
	\label{table:indexes}
	\begin{tabular}{l c l c} 
	\hline
	\hline
    Stage  & Time Interval(s) & Parameter & Value \\
	\hline
	Rise        &   $600 \mbox{s} < T < t_{p}$ &     $\alpha_{O,\rm Rise}$     &  $1.53 \pm 0.06$    \\
    Peak        &    $t_{\rm peak} \approx 1450$ s  &                           &                    \\
    Early decay &  $t_{\rm peak} < T < 5000$ s & $\alpha_{O,\rm Early}$ & $-1.41 \pm 0.02$   \\
    Plateau     & $5000~\mbox{s} < T < 11000$ &    $\alpha_{O,\rm Plateau}$   & $-0.65 \pm 0.07$   \\
    Late decay  &    $11000 < T < 21500 $    & $\alpha_{O,\rm Late}$  & $-1.36 \pm 0.04$   \\
    Jet break   &   $21000 < T < 45000 $    & $\alpha_{O,\rm Break}$  & $-1.94 \pm 0.08$   \\
    \hline
	\hline
	\end{tabular}
\end{table}


\subsection{Redshift}\label{sec:Redshift}

We used the  RATIR afterglow photometry in $griZYJH$ to build the spectral energy distribution (SED) of GRB 191016A. Since the afterglow is variable, we determined the mean magnitudes in the interval $T+6000$~s to $T+9000$~s, during which we have uniform coverage with these filters. The SED is shown in Table~\ref{tab:SED-photometry}. 

To determine the photometric redshift $z$, we applied the fitting algorithm developed by \cite{2014AJ....148....2L}, which we describe briefly here. We assume that for the regime in which RATIR observes the intrinsic shape of the SED is described with a single power-law: $F(\lambda)=F_{0}\lambda^{\beta}$, where $F(\lambda)$ is the flux density as a function of wavelength. To account for host galaxy dust extinction we use three standard templates for the extinction law (Milky Way, Small Magellanic Cloud and Large Magellanic Cloud) from \cite{1992ApJ...395..130P}. We also considered one fit where no host extinction is assumed ($A_V=0$). We included the attenuation from the intervening intergalactic medium (IGM) in our modelling by estimating the reduction of flux from Ly$\alpha$, Ly$\beta$, and Ly-continuum absorption using a similar methodology to that in the {\sc hyperz} software (see also \citealp{2000A&A...363..476B}). However, we do not consider damped Ly$\alpha$ (DLA) absorption associated with the host galaxy.

\begin{figure}
	\includegraphics[trim={0 0 0 1.4cm},clip,width=\columnwidth]{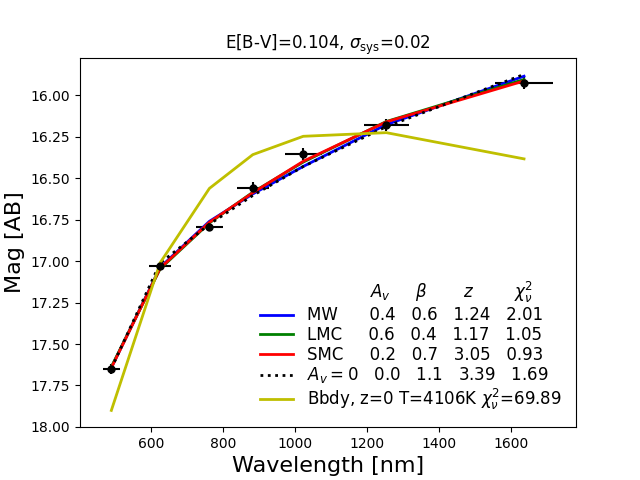}
    \caption{Redshift determination. {\itshape Top:} Spectral energy distribution from the RATIR photometry. Magnitudes in the plot are corrected for Galactic extinction in the source direction using E[B-V] = 0.104. Fitted solutions are presented for the different host extinction laws considered: MW (solid blue line), SMC (solid red line), LMC (solid green line) or A$_{v}$=0 (dashed black line). Our best-fitting model yields a redshift value between $2.47 - 3.33$ for a reduced $\chi^{2}_{\nu}$ = 0.93, at a 90$\%$ confidence level. The extinction value A$_{v}$, spectral optical index $\beta$, redshift $z$ and reduce $\chi^{2}_{\nu}$ obtained from each model are also shown for reference. }
    \label{fig:redshift}
\end{figure}

We found that the best $\chi^{2}_{\nu}$ value is achieved for using a SMC extinction law, with $A_V\approx0.2$ and $\beta \approx -0.7$, and $z$ in the range of $2.47 - 3.33$. Our best-fitting model is shown in Figure \ref{fig:redshift}, where the results from other models are also quoted for reference. 

\begin{table}
	\centering
	\caption{The SED of GRB 191016A from RATIR observations between $T+6000$~s and $T+9000$~s.} 
	\label{tab:SED-photometry}
	\begin{tabular}{cccc} 
	\hline
	\hline
Filter & Wavelength (nm) & AB \\
    \hline
    $g$  & \phantom{0}489  & $18.01 \pm 0.02$ \\
    $r$  &  \phantom{0}625  & $17.29 \pm 0.01$ \\
    $i$  &  \phantom{0}762  & $16.99 \pm 0.01$ \\ 
    $Z$  &  \phantom{0}883  & $16.71 \pm 0.02$ \\ 
    $Y$  & 1023 & $16.48 \pm 0.01$ \\ 
    $J$  & 1254 & $16.26 \pm 0.01$ \\ 
    $H$  & 1637 & $15.98 \pm 0.01$ \\ 
	\hline    
     \end{tabular}
\end{table}

To constraint the spectral optical index $\beta_{o}$ we interpolated our late time observations from RATIR-$\emph{r}$ and RATIR-$\emph{i}$ bands to the times of XRT observations, between $T+45000$~s and $T+56900$~s. We found a spectral slope from optical to X-ray emission of $0.68\pm0.08$,in agreement with the beta value for the SMC model in our fittings. 

\cite{2021ApJ...911...43S} also estimated the redshift of GRB 191016A from the GROND $grizJHK$ SED \citep{2019GCN.26176....1S}. They find $z=3.29\pm0.4$, which lies well inside the redshift range from our fits. We note four differences in our analyses. First, the uncertainties in our RATIR SED are 1-2\% and as such are much lower than the 3-8\% uncertainties in the GROND SED. Second, \cite{2021ApJ...911...43S} include host damped Ly$\alpha$ absorption whereas we do not.
Finally, the GROND SED has $K$ but not $Y$ whereas the RATIR SED has $Y$ but not $K$. That said, the uncertainty in the GROND measurement of $K$ is 8\%, so it cannot provide an especially strong constraint on the redshift. Finally, we note that the reduced $\chi^2_{\nu}$ for the \cite{2021ApJ...911...43S} fit is 4.1, whereas the reduced $\chi^2_{\nu}$ for our fit is 0.93 (despite our more precise photometry). 

Throughout this work, we will refer to our redshift estimated range $2.47 \leq z  \leq 3.33$ as the most likely redshift for GRB 191016A and will use $z=3.29\pm0.4$ in our calculations for consistency with \cite{2021ApJ...911...43S}. 


\section{DISCUSSION}\label{sec:Discussion}

Our analysis suggests that the temporal evolution of the optical afterglow of GRB 191016A can be described in 5 different stages, shown in Figure \ref{fig:PowerLaw_Fits}, that were not completely characterized in previous works. Based on TESS Sector 17 observations, \cite{2021ApJ...911...43S} successfully identify the rise and decay phases of the optical afterglow but failed in determining the exact peak time, mainly because of the sparse sampling in their data. The 10 to 30 minutes TESS cadence also dismiss the plateau phase that is clearly detected in our RATIR and COATLI light curves, between $T+5000$~s to $T+11000$~s. This plateau phase has been recently reported by \cite{2021arXiv211109123S}, whose analysis also suggests a probable jet break occurring at later times. Based on a larger quantity of photometric data, our results support the existence of a jet break with a temporal index $\alpha = 1.94$, in excellent agreement with the temporal slope obtained from X-ray observations. In the following sections we discuss our results in the statistical context, comparing the properties of GRB 191016A afterglow with a larger sample of GRBs with similar light curves, and propose a theoretical interpretation of the observed behaviour in the optical RATIR and COATLI data.

\subsection{GRB 191016A in a statistical context} \label{sec:Context}
\begin{figure*}
\centering
    \includegraphics[width=\columnwidth]{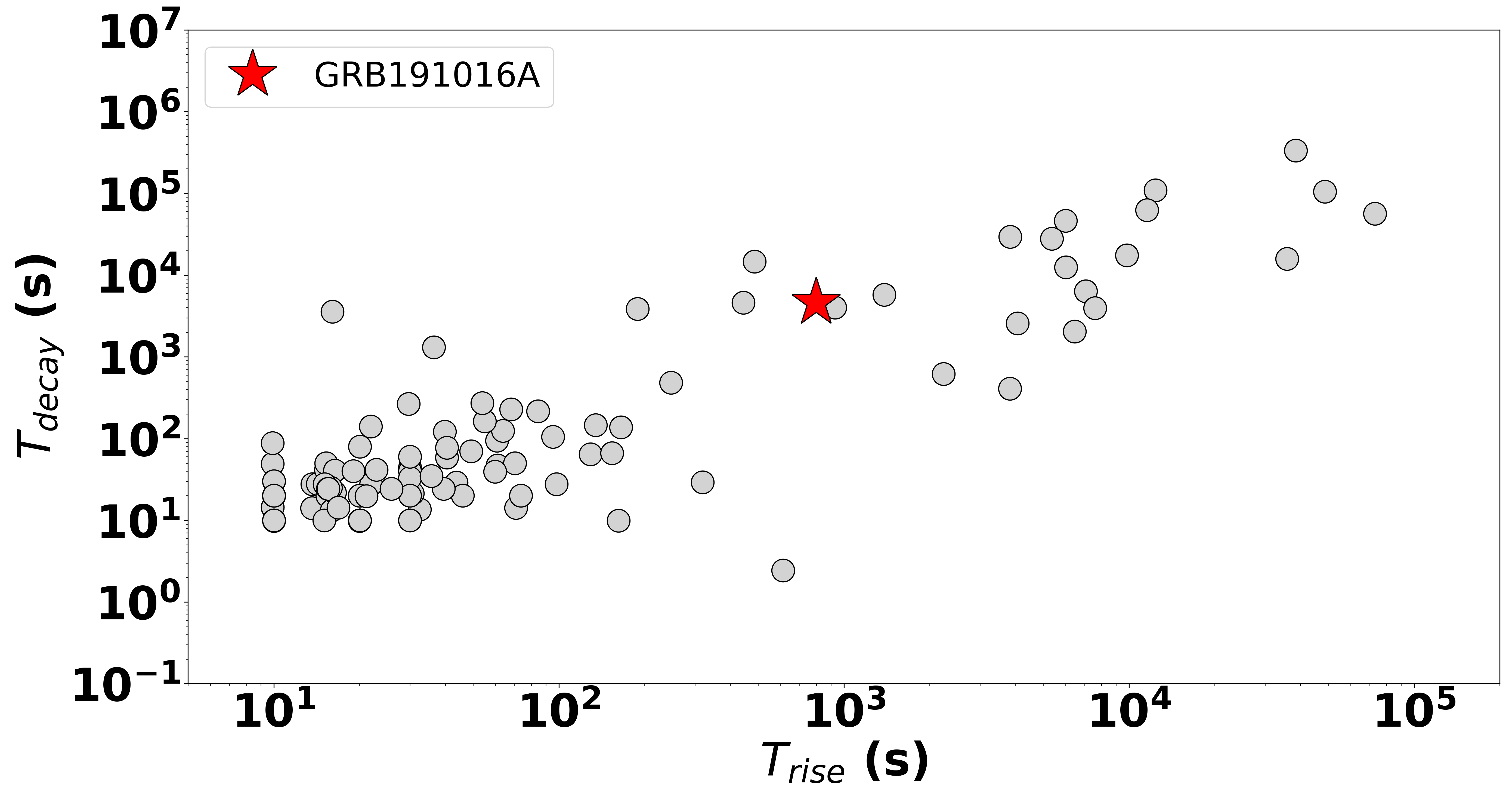}
    \includegraphics[width=\columnwidth]{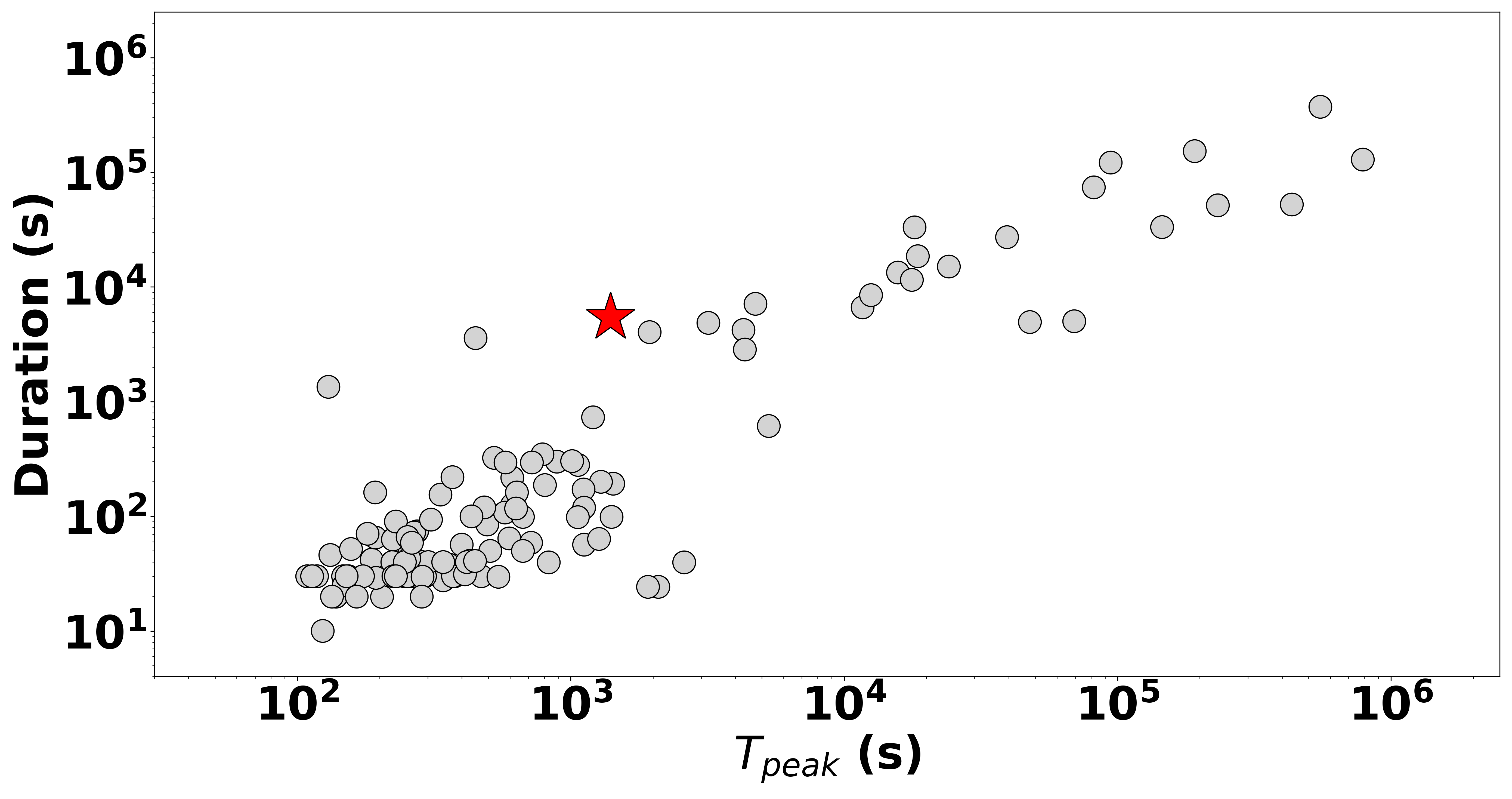}
    \includegraphics[width=\columnwidth]{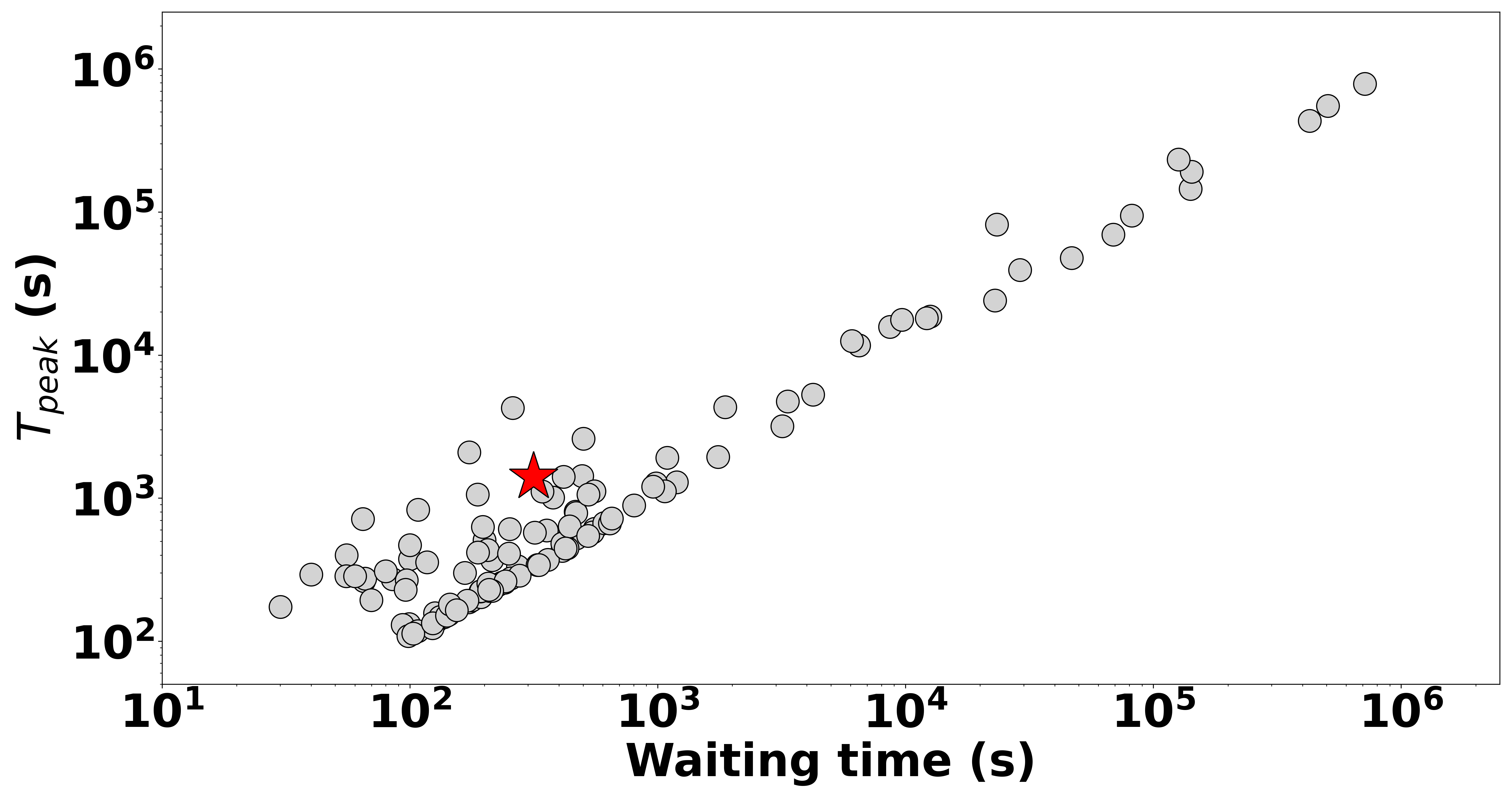}
    \includegraphics[width=\columnwidth]{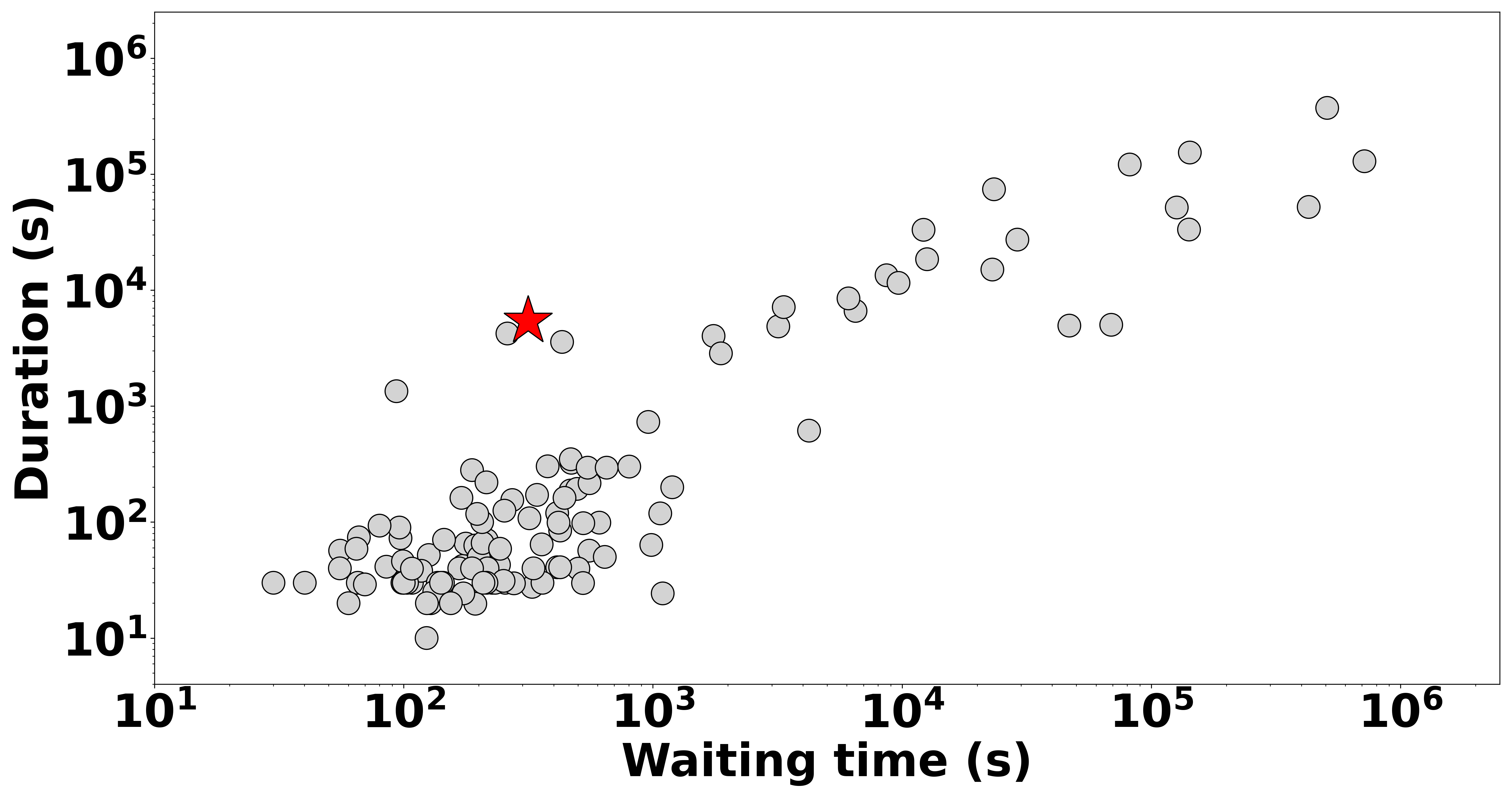}
    \caption{Correlations between timescales of GRB optical flares (taken from \citealt{2017ApJ...844...79Y}). The red star represents the location of GRB 191016A within the sample of GRB where optical flares has been observed (in light grey). The rise time is correlated with the decay time, and the duration is correlated with the peak time, which are both consistent with the results of the X-ray flares.}
    \label{fig:Statistics}
\end{figure*}

In the framework of the standard forward shock model \citep{1992MNRAS.258P..41R,1997ApJ...476..232M,1998ApJ...503..314P,1998ApJ...497L..17S}, the onset of the GRB afterglow is characterized by a smooth optical rising light curve when the ultra-relativistic fireball is decelerated by the circumburst medium which is expected to be seen at the early stages of the afterglow emission. \citet{2009MNRAS.395..490O} showed that, in the observer's frame, a significant fraction of the optical afterglow light curves rise in the first 500s after the GRB trigger and decay after. Our observations show clear evidence for a rising afterglow in the light curve of GRB 191016A, but it peaks a relatively late times, around $t_{\rm peak}\sim 1450$~s. We further investigated this by comparing the timescales of the optical bump we detected in GRB 191016A with the statistical distributions of the optical flares studied by \cite{2017ApJ...844...79Y}, presented in Figure \ref{fig:Statistics}. Their conclusions on the observed correlations between duration, rise and peak time can be summarized as follows: \emph{i)} longer rise times are associated with longer decay times, suggesting that broader optical flares peak at later times, \emph{ii)} waiting time is also correlated with duration time of optical flares and peak time, showing that a longer waiting time corresponds to a broader optical flare that peaks at a later time. These results are consistent with our findings for the timescales observed in the GRB 191016A: the optical flare is broad, occurring after a long waiting time, corresponding to a long rise and long decay, and it peaks at late times.
Although at first glance a late peaking afterglow might seem atypical, previous works by \cite{2018A&A...609A.112G} and \cite{2010ApJ...720.1513K} have proven that GRB afterglows with peak times of $10^{3}$ are not unprecedented. A brief discussion on some other physical scenarios to explain the late peak observed in GRB 191016A afterglow can be found in \cite{2021ApJ...911...43S} and \cite{2021arXiv211109123S}.

\subsection{Theoretical Interpretation}\label{sec:Interpretation} 

\subsubsection{The bright peak followed by the normal decay}
A bright peak observed in the optical light curve is usually associated with the onset of the afterglow \citep[e.g., see][]{2007A&A...469L..13M}.  It is widely accepted that the optical peak followed by the normal decay phase is described by the standard synchrotron emission from forward shock (FS) or reverse shock (RS), or a superposition of both components evolving in a uniform-density medium \citep{1998ApJ...497L..17S, 2003ApJ...582L..75K, 2016ApJ...818..190F, 2019ApJ...887..254B, 2019ApJ...881...12B}. The monotonic rise and decay in the GRB 191016A light curve, has been previously observed in the afterglow of a small sample of GRBs by \cite{2017ApJ...845...52L} and is usually attributed to a single forward shock wave. However, the universal scaling function proposed to describe GRBs afterglow behaviour based on this assumption requires a rising temporal index $\alpha$ = 2, whereas the statistical results presented by \citet{2010ApJ...725.2209L} show that the rising index is mostly between 1 and 2 for the majority of GRBs where the onset of the optical afterglow is clearly detected. With a rising temporal index $\alpha$ = 1.53, we conclude that the afterglow evolution of GRB 191016A is quite consistent with the general behaviour observed in GRB's with rising afterglows but cannot be explained by an evolving FS only. 
Considering the framework of the RS models for an ISM scenario, evolving in a thin-shell case (i.e. a RS crossing time larger than GRB duration, as in the case of GRB 191016A), when the RS emission dominates over the FS a rising slope between $3-5$ with a steep decay temporal index around $-2$ are expected \citep{2016ApJ...833..100H,2020ApJ...895...94Y}. Given the temporal index found in the optical afterglow of GRB 191016A, we conclude that its light curve behaviour cannot be explained with models where only RS emission is assumed. On the other hand, it is also worth noting that the initial rise is too steep to be interpreted by the synchrotron RS scenario from a stellar wind medium \citep[e.g., see][]{2013NewAR..57..141G}.

The synchrotron light curves in each region depend on the cooling regime, the values of synchrotron spectral breaks, the maximum flux and the evolution of the bulk Lorentz factor during the coasting and deceleration phase. In the FS region, initially the synchrotron flux in the fast (slow)-cooling regime evolves as $F_{\rm \nu, f} \propto t^2 (t^3)$ for  $\nu_{\rm c, f}(\nu_{\rm m, f})<\nu_\gamma<\nu_{\rm m, f}(\nu_{\rm c, f})$ and $\propto t^2 (t^2)$ for  $\nu_{\rm m, f}(\nu_{\rm c, f}) <\nu_\gamma$ \citep{2013NewAR..57..141G}, and later it evolves as $F_{\rm \nu, f} \propto t^{-\frac{1}{4}} (t^{-\frac{3p-3}{4}})\,$ for $\nu_{\rm c, f}(\nu_{\rm m, f})<\nu_\gamma<\nu_{\rm m, f}(\nu_{\rm c, f})$ and $\propto t^{-\frac{3p-2}{4}}(t^{-\frac{3p-2}{4}})$ for $\nu_{\rm m, f}(\nu_{\rm c, f})<\nu_\gamma$, respectively, with $p$ the spectral index of the electron distribution (see  \citealp{1998ApJ...497L..17S} for further details).  In the RS region, before the shock crossing time the synchrotron flux in the fast (slow)-cooling regime evolves as  $F_{\rm \nu, r}\propto t^{\frac{1}{2}}(t^{\frac{6p-3}{2}})$ for  $\nu_{\rm c, r}(\nu_{\rm m, r})<\nu_\gamma<\nu_{\rm m, r}(\nu_{\rm c, r})$ and $\propto t^{\frac{6p - 5}{2}}(t^{\frac{6p - 5}{2}})$ for $\nu_{\rm m, r}(\nu_{\rm c, r})<\nu_\gamma$, and after the shock crossing time it evolves as $F_{\rm \nu, r} \propto  t^{-\frac{16}{35}}$ for $\nu_\gamma<\nu_{\rm m, r}$ and $t^{-\frac{27p+7}{35}}$ for $\nu_{\rm m, r}<\nu_\gamma<\nu_{\rm cut}$ \citep{2013NewAR..57..141G,2000ApJ...545..807K}. The terms $\nu_{\rm m}$, $\nu_{\rm c}$ and $\nu_{\rm cut}$ are the characteristic,  cooling and cut-off spectral breaks, respectively (as described by \citealp{2013NewAR..57..141G, 1998ApJ...497L..17S, 2000ApJ...545..807K}). 
Therefore, in this work we used a superposition of synchrotron light curves from  FS and RS regions evolving in a thin-shell case and constant medium to describe the bright peak flare observed in GRB 191016A.  Before the bright peak, the optical flux evolving with temporal index of $\alpha=1.53\pm0.06$ is consistent with a superposition of synchrotron FS and RS model, and after the bright peak, the optical flux with temporal decay of $\alpha=-1.41\pm0.03$ is consistent with the synchrotron FS model evolving in the power-law segment $\nu_{\rm c, f}< \nu$ for $p=2.55\pm0.04$.

\subsubsection{Late Energy injection into the afterglow blast wave}
To better explain the optical excess observed in the plateau phase, from $ T + 5000\,{\rm s}$ to $T + 11000\,{\rm s}$, we must consider a late energy injection scenario. Late energy injection by the central engine on the GRB afterglow can produce refreshed shocks. We assume that the central engine is a black hole (BH), and the fallback accretion could trigger the energy extraction from the BH via Blandford-Znajek (BZ) mechanism.  The BZ jet power from a BH with mass $M_{\rm BH}$ and angular momentum $J$ can be described as
\begin{equation}\label{L_BZ}
    L_{\rm BZ}=9.3\times 10^{53}{\rm erg\, s^{-1}}\frac{a^2 \dot{m} F(a)}{(1+\sqrt{1-a^2})^2}\,,
\end{equation}
where $a=Jc/G M^2_{\rm BH}$ is the dimensionless spin parameter, $F(a)=[(1+s^2)/s^2][(s+1/s)\arctan s - 1]$  with $s=a/(1+\sqrt{1-a^2})$,  $\dot{m}=\frac{\dot{M}}{M_\odot\,s^{-1}}$ is the dimensionless accretion rate onto the BH with $\dot{M}$ given by  \citep{2008Sci...321..376K, 2008MNRAS.388.1729K}
\begin{equation}\label{M_odot}
    \dot{M}=\frac{1}{\tau_{\rm vis}}e^{-\frac{t}{\tau_{\rm vis}}}\int^t_{t_0}e^{\frac{t'}{\tau_{\rm vis}}}\dot{M}_{\rm fb}dt'
\end{equation}
and $\dot{M}_{\rm fb}$ the fallback accretion rate described by \citep{2000ApJ...545..807K, 2001ApJ...550..410M}

\begin{equation}\label{M_fb}
    \dot{M}_{\rm fb}= \dot{M}_{\rm p}\left[\frac12\left(\frac{t-t_0}{t_{\rm p}-t_0}\right)^{-1/2} +\frac12\left(\frac{t-t_0}{t_{\rm p} - t_0}\right)^{5/3} \right]^{-1}\,. 
\end{equation}
The terms $\tau_{\rm vis}$ is the viscous timescale, $\dot{M}_{\rm p}$ is the peak fallback rate, $t_0$ and $t_{\rm p}$ are the start and peak time of the fallback accretion, respectively.\\

From Eqs. \ref{L_BZ}, \ref{M_odot} and \ref{M_fb}, one can see that the energy injected into the blast wave corresponds to $E_{\rm inj}(t)=\int L_{\rm BZ}dt$, which depends on evolution of the fallback accretion rate. In this case,  during the deceleration phase the synchrotron FS light curves with continuous energy injection for the fast(slow)-cooling regime in a uniform-density medium are obtained following \cite{2006ApJ...642..354Z}.

{\rm The temporal decay indexes observed in the X-ray and optical bands for $4\times 10^4\,{\rm s}\lesssim t$ are similar to each other and are steeper than the temporal indexes of the normal decay phase. Both indexes are consistent with the post jet-break decay phase with energy injection for $p=2.55$ and $L_{\rm BZ}\propto t^{-q}$ with an energy injection index of $q\approx 0.85$} \citep{2018ApJ...859..160W}.

\subsubsection{Hydrodynamical evolution}

The BZ mechanism injects energy continuously into the blast wave increasing the bulk Lorentz factor, which causes a rise in the observed flux. We consider the dynamical equations proposed by  \cite{1999MNRAS.309..513H, 2000MNRAS.316..943H}. The dynamical evolution of the outflow into the circumburst medium  can be described  by

\begin{eqnarray}
\frac{dr}{dt}&=&\beta c \Gamma \left(\Gamma + \sqrt{\Gamma^2-1}\right)\cr
\frac{dm}{dr}&=&2\pi\,(1-\cos\theta)r^2 n m_p\cr
\frac{d\theta}{dt}&=&\frac{c_s}{r}(\Gamma + \sqrt{\Gamma^2-1})\cr
\frac{d\Gamma}{dm}&=&-\frac{\Gamma^2 -1 - \frac{L_{\rm BZ}}{c^2}\frac{dt}{dm}}{M_{\rm ej}+ 2\Gamma m}\,,
\end{eqnarray}

where $c_s=\sqrt{\hat\gamma(\gamma-1)(\Gamma-1)c^2/\left(1 + \hat\gamma(\Gamma -1)\right)}$ with $\hat\gamma\approx (4\Gamma + 1)/3\Gamma$,  $M_{\rm ej}=E_0/\Gamma_0c^2$ is the initial value of the ejected mass with $\Gamma_0$ the initial Lorentz factor and $m$ is the swept-up mass by the shock. The term $E_0$ is  the initial energy per solid angle $E_{\rm iso} \Omega_0/4\pi$ with 
$\Omega_0=2\pi(1-\cos\theta)$ and $\theta$ the half-opening angle of the relativistic jet. It is worth noting that the jet break transition takes place when the Lorentz factor becomes inversely proportional to the half-opening angle ($\Gamma \approx 1/\theta$). The previous equations are consistent with the self-similar solution during the ultra-relativistic phase \citep[Blandford-McKee solution;][]{1976PhFl...19.1130B}. The dynamics and radiation equations of the synchrotron model from FS and RS derived in \cite{1998ApJ...497L..17S} and \cite{2000ApJ...545..807K}, respectively, are used.


\subsubsection{Modelling GRB 191016A afterglow light curve}
Based upon our estimated redshift for the optical afterglow of GRB 191016A, in our proposed model we used a value of $z=3.29$ and assumed a typical isotropic-equivalent energy $E_{\rm iso}=2.5\times 10^{54}\,{\rm erg}$. We also assumed a constant density of $n=0.8\,{\rm cm^{-3}}$ for the circumburst medium. Since the optical observations evolving with a decay index of $-0.65\pm0.07$ between $ T + 5000\,{\rm s}$ and $T + 11000\,{\rm s}$ are not consistent with the standard synchrotron FS and RS model, we also included a late energy injection phase.\\ 

\begin{figure}
\centering
    \includegraphics[width=\columnwidth]{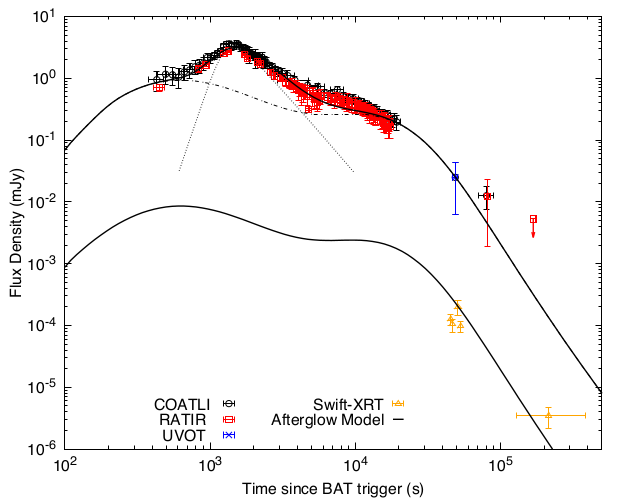}
    \caption{\textbf{Theoretical interpretation of the optical afterglow.} We present our proposed model for the afterglow emission of the GRB 191016A, superimposed over the Coatli-{\itshape w} (black circles), RATIR-{\itshape r} (red squares) and UVOT-{\itshape w} (blue star) data. The different components of the afterglow are shown independently using dotted and dashed lines. The bright peak flare is described by synchrotron emission from FS (dashed line) with an embedded RS (dotted line), and the small bump observed at $\sim 10^4$~s by synchrotron emission from a late energy injection into the blast wave. We also plot late time X-ray observations to illustrate the consistency of our model, particularly for the temporal slope obtained from the optical data and high-energy data in the jet break phase.}
    \label{fig:TheoreticalModel}
\end{figure}

We show in Figure \ref{fig:TheoreticalModel} the proposed model for the optical afterglow observations using a superposition of synchrotron light curves from  FS and RS evolving in a thin-shell case and constant medium.   The bright peak flare is described by synchrotron emission from  FS and RS, and the small bump observed at $\sim 10^4$~s by synchrotron emission from refreshed shocks. These shocks are consistent with late energy injection into the afterglow blast wave, obtained by fallback accretion onto the BH. A substantial contribution of the synchrotron emission from the reverse shock is clearly seen around $t_{\rm peak}\sim 1.4\times 10^3\,{\rm s}$, producing a small bump over the FS emission at the peak time. \\

Table \ref{tab:fit_par} shows the values of parameters derived in our proposed model The ratio of the microphysical parameters in the forward and reverse-shock region $\mathcal{R}=\epsilon_{\rm B,r}/\epsilon_{\rm B,f}\approx 7$ indicates that the outflow is magnetized. This model is also supported by the evidence of polarization at the beginning of this plateau phase reported by \cite{2021arXiv211109123S}. It suggests that the outflow must have dissipated part of the Poynting flux during the prompt phase. The value of constant density $n=0.8\,{\rm cm^{-3}}$ gives an initial bulk Lorentz factor $\Gamma = 135$ that lies within the tail of the cumulative distribution reported by \cite{2018A&A...609A.112G}, for bursts exhibiting a bright peak and leads to a deceleration radius of $\approx 4\times 10^{17}\,{\rm cm}$.

We have applied the fall-back accretion process to explain the small optical bump and have required the BZ mechanism by considering the central object is a black hole. The values obtained after we describe the small optical bump corresponds to the typical values found for other burst \citep{2015MNRAS.446.3642Y, 2016ApJ...826..141G}.   The fall-back accretion rate reaches the peak value of $2.7\times 10^{-9}\,{\rm M_{\odot}s^{-1}}$ at $1.9\times 10^4\,{\rm s}$.   The fall-back material is assumed to be continuous, although clumps might sometimes appear in the material and increase dramatically the fall-back accretion rate.   The viscous timescale of $4.0\times 10^3\,{\rm s}$ is relatively small, inferring that fall-back accretion lies in the fast accretion regime. Given the value of the start time of the fall-back accretion ($10^3\,{\rm s}$) and assuming a typical BH mass of $2.2 M_\odot$, the minimum radius of the fall-back material becomes $\sim 0.8\times 10^{11}\,{\rm cm}\,\left( \frac{M_{\rm BH}}{2.2 M_\odot}\right)^\frac13 \left( \frac{t_0}{10^3\,{\rm s}}\right)^{\frac23}$, which corresponds to the radius of a Wolf-Rayet star.\\

From a theoretical perspective, different models have been used to explain the different temporal indexes found in the afterglow light curves of GRB with optical peaks. For instance, the inclusion of a stratified medium has been proposed by some authors for the FS models \citep{2013ApJ...776..120Y}, concluding that the circumburst environment might not be either a homogeneous ISM or a typical stellar wind but something in between, with a general density distribution of $n \propto  r^{-k}$ for $k$ values in the range of 0.4-1.4. For $k$ in general, the evolution of bulk Lorentz factors $\Gamma\propto t^0$ and $\Gamma\propto t^{\frac{k-3}{8-2k}}$ before and after the deceleration phase, respectively, imply that the synchrotron flux in the fast(slow)-cooling regime evolves as $F_{\rm \nu, f} \propto t^{\frac{8-3k}{4}} (t^{\frac{12-k(p+5)}{4}})$ for  $\nu_{\rm c, f}(\nu_{\rm m, f})<\nu_\gamma<\nu_{\rm m, f}(\nu_{\rm c, f})$ and $\propto t^{\frac{8-k(p+2)}{4}} (t^{\frac{8-k(p+2)}{4}})$ for  $\nu_{\rm m, f}(\nu_{\rm c, f}) <\nu_\gamma$ before the deceleration phase, and  as $F_{\rm \nu, f} \propto t^{-\frac{1}{4}} (t^{-\frac{12(p-1) + k(5-3p)}{4(4-k)}})\,$ for $\nu_{\rm c, f}(\nu_{\rm m, f})<\nu_\gamma<\nu_{\rm m, f}(\nu_{\rm c, f})$ and $\propto t^{-\frac{3p-2}{4}}(t^{-\frac{3p-2}{4}})$ for $\nu_{\rm m, f}(\nu_{\rm c, f})<\nu_\gamma$, after the deceleration phase.  For GRB 191016A, the value of  $k=0.42$ is consistent with the evolution of temporal rise and decay indexes of synchrotron flux in the cooling condition $\nu_{\rm m, f}< \nu_{\rm c, f}<\nu_\gamma$. However, a transition from stratified to homogeneous medium with identical temporal index is not observed.


\begin{table}
\caption{Parameters found of the early and late afterglow of GRB 191016A \label{tab:fit_par}}
\centering
\begin{tabular}{lcc}\hline\hline\\
Parameter& Symbol& Value\\
\hline\hline\\
electron PL index& $p$ & $2.52$\,\\
ISM density & $n$ & $0.8$\,(cm$^{-3}$)\,\\
Initial Lorentz Factor& $\Gamma_0$&135\\

&$\epsilon_{\rm B,f}$& $3\times 10^{-5}$\\
&$\epsilon_{\rm B,r}$&$4\times 10^{-4}$\\
&$\epsilon_e$&$5\times 10^{-2}$\\
Isotropic energy&$E_{\rm iso}$ (erg)&$2.5\times 10^{54}$ \\
Half-opening angle&$\theta$ (degree)&$3.7$ \\
Redshift &z&$3.29$\\


\hline \\
Spin parameter& $a$ &0.9\,\\
Start time& $t_{0}$ &$10^3\,({\rm s})$\,\\
Peak time& $t_{\rm p}$ & $1.9\times 10^4\,({\rm s})$ \\
Viscous timescale & $\tau_{\rm vis}$ &$2.0\times 10^3\,({\rm s})$ \\
Peak fallback rate & $\dot{M}_{\rm p}$ & $2.7\times 10^{-9}\,(M_{\odot}\,s^{-1})$\\ 
\hline\hline\\
\end{tabular}
\label{table1}
\end{table}

\section{Conclusions} \label{sec:Conclusions}

We present the most complete temporal coverage existing to date for the optical emission associated to the GRB 191016A. We describe its temporal evolution in five different stages: {\itshape rise}, {\itshape early decay}, {\itshape plateau}, {\itshape late decay} and {\itshape a jet break} Our observations complement previous works in which the peak time, the plateau phase and the subsequent late decay followed by the jet break, were not well characterized for this GRB. We report a peak time for the afterglow emission of around $1450 \pm 50$~s and a temporal index $\alpha = 1.94$ for the jet break, occurring between 6 and 12 hrs after the BAT trigger, in excellent agreement with X-ray observations at late times. 

Based on RATIR photometry, we provide additional constraints on the redshift of the GRB191016A. We place the source at high redshift,  consistent with previous works, in the range of $2.47 \leq z  \leq 3.33 $. In a statistical context, we conclude that the relatively late time of the optical peak is not particularly exceptional when comparing with the whole population of GRB with rising afterglows. 

We propose a theoretical interpretation for the afterglow of GRB 191016A in terms of a superposition of standard synchrotron light curves from  forward- and reverse-shock regions. The optical observations are consistent with an afterglow expanding in a constant density  medium and a reverse shock evolving in a thin-shell regime.  The bright peak flare is described by synchrotron emission from FS and RS, and the small bump observed at $\sim 10^4$~s by synchrotron emission from refreshed shocks. These shocks are consistent with late central activity, obtained by fallback accretion onto the BH, providing energy injection into the afterglow blast wave at later times.  


\section*{Acknowledgements}

Some of the data used in this paper were acquired with the RATIR instrument, funded by the University of California and NASA Goddard Space Flight Center, and the 1.5-meter Harold L. Johnson telescope at the Observatorio Astronómico Nacional on the Sierra de San Pedro Mártir, operated and maintained by the Observatorio Astronómico Nacional and the Instituto de Astronomía of the Universidad Nacional Autónoma de México. Operations are partially funded by the Universidad Nacional Autónoma de México (DGAPA/PAPIIT IG100414, IT102715, AG100317, IN109418, IG100820, and IN105921). We acknowledge the contribution of Leonid Georgiev and Neil Gehrels to the development of RATIR.

Some of the data used in this paper were acquired with the DDOTI instrument and the COATLI telescope and interim instrument at the Observatorio Astronómico Nacional on the Sierra de San Pedro Mártir. DDOTI and COATLI are partially funded by CONACyT (LN 232649, LN 260369, LN 271117, and 277901), the Universidad Nacional Autónoma de México (CIC and DGAPA/PAPIIT IG100414, IT102715, AG100317, IN109418, and IN105921), the NASA Goddard Space Flight Center. DDOTI is partially funded by the University of Maryland (NNX17AK54G). DDOTI and COATLI are operated and maintained by the Observatorio Astronómico Nacional and the Instituto de Astronomía of the Universidad Nacional Autónoma de México. We acknowledge the contribution of Neil Gehrels to the development of DDOTI. RLB acknowledges support from the DGAPA-UNAM and CONACyT postdoctoral fellowships.

We thank the staff of the Observatorio Astronómico Nacional.

This work made use of data supplied by the UK Swift Science Data Centre at the University of Leicester. 

\section*{Data Availability}
Optical and NIR photometry used in this work is presented in the Appendix A. The BAT, XRT and UVOT data sets underlying this article are available in the domain https://www.swift.ac.uk/xrt$\_$products/index.php.


\bibliographystyle{mnras}


\appendix 

\section{Photometry used in this work} \label{Tables_photometry}

\clearpage
\begin{table}
	\centering
	\caption{COATLI binned optical photometry for GRB 191016A.} 
	\label{tab:COATLI-photometry_01}
	\begin{tabular}{cccc} 
	\hline
$t_{i}$ (s) & $t_{f}$ (s) & $t_{exp}$ (s) & \emph{w} (mag) \\ 
	\hline 
430.0 & 460.0 & 30.0 & 16.26 $\pm$ 0.06 \\ 
494.0 & 524.0 & 30.0 & 16.04 $\pm$ 0.05 \\ 
552.0 & 582.0 & 30.0 & 16.06 $\pm$ 0.06 \\ 
609.0 & 639.0 & 30.0 & 16.15 $\pm$ 0.06 \\ 
665.0 & 695.0 & 30.0 & 15.94 $\pm$ 0.05 \\ 
724.0 & 754.0 & 30.0 & 15.86 $\pm$ 0.05 \\ 
788.0 & 818.0 & 30.0 & 15.74 $\pm$ 0.04 \\ 
844.0 & 874.0 & 30.0 & 15.68 $\pm$ 0.05 \\ 
899.0 & 929.0 & 30.0 & 15.59 $\pm$ 0.04 \\ 
956.0 & 986.0 & 30.0 & 15.37 $\pm$ 0.03 \\ 
1012.0 & 1042.0 & 30.0 & 15.37 $\pm$ 0.03 \\ 
1069.0 & 1099.0 & 30.0 & 15.28 $\pm$ 0.02 \\ 
1125.0 & 1155.0 & 30.0 & 15.22 $\pm$ 0.02 \\ 
1181.0 & 1211.0 & 30.0 & 15.12 $\pm$ 0.02 \\ 
1239.0 & 1269.0 & 30.0 & 14.94 $\pm$ 0.02 \\ 
1296.0 & 1326.0 & 30.0 & 14.83 $\pm$ 0.02 \\ 
1353.0 & 1383.0 & 30.0 & 14.80 $\pm$ 0.02 \\ 
1409.0 & 1439.0 & 30.0 & 14.90 $\pm$ 0.02 \\ 
1465.0 & 1495.0 & 30.0 & 14.90 $\pm$ 0.02 \\ 
1521.0 & 1551.0 & 30.0 & 14.81 $\pm$ 0.02 \\ 
1577.0 & 1607.0 & 30.0 & 14.84 $\pm$ 0.02 \\ 
1634.0 & 1664.0 & 30.0 & 14.95 $\pm$ 0.02 \\ 
1690.0 & 1720.0 & 30.0 & 14.96 $\pm$ 0.02 \\ 
1746.0 & 1776.0 & 30.0 & 15.01 $\pm$ 0.02 \\ 
1803.0 & 1833.0 & 30.0 & 15.10 $\pm$ 0.03 \\ 
1858.0 & 1888.0 & 30.0 & 15.24 $\pm$ 0.05 \\ 
1915.0 & 1945.0 & 30.0 & 15.20 $\pm$ 0.02 \\ 
1971.0 & 2001.0 & 30.0 & 15.23 $\pm$ 0.02 \\ 
2028.0 & 2058.0 & 30.0 & 15.32 $\pm$ 0.03 \\ 
2060.0 & 2065.0 & 5.0 & 15.35 $\pm$ 0.06 \\ 
2073.0 & 2103.0 & 30.0 & 15.33 $\pm$ 0.02 \\ 
2144.0 & 2174.0 & 30.0 & 15.30 $\pm$ 0.03 \\ 
2178.0 & 2208.0 & 30.0 & 15.43 $\pm$ 0.03 \\ 
2215.0 & 2245.0 & 30.0 & 15.43 $\pm$ 0.03 \\ 
2285.0 & 2315.0 & 30.0 & 15.36 $\pm$ 0.04 \\ 
2356.0 & 2386.0 & 30.0 & 15.52 $\pm$ 0.04 \\ 
2390.0 & 2420.0 & 30.0 & 15.56 $\pm$ 0.03 \\ 
2427.0 & 2457.0 & 30.0 & 15.54 $\pm$ 0.03 \\ 
2550.0 & 2670.0 & 120.0 & 15.67 $\pm$ 0.02 \\ 
2692.0 & 2812.0 & 120.0 & 15.77 $\pm$ 0.02 \\ 
2834.0 & 2954.0 & 120.0 & 15.86 $\pm$ 0.03 \\ 
2975.0 & 3095.0 & 120.0 & 15.93 $\pm$ 0.03 \\ 
3151.0 & 3271.0 & 120.0 & 16.00 $\pm$ 0.03 \\ 
3431.0 & 3551.0 & 120.0 & 16.10 $\pm$ 0.03 \\ 
3836.0 & 4316.0 & 480.0 & 16.28 $\pm$ 0.03 \\ 
4692.0 & 5172.0 & 480.0 & 16.48 $\pm$ 0.03 \\ 
5502.0 & 5982.0 & 480.0 & 16.54 $\pm$ 0.03 \\ 
6127.0 & 6607.0 & 480.0 & 16.79 $\pm$ 0.03 \\ 
6744.0 & 7224.0 & 480.0 & 16.83 $\pm$ 0.03 \\ 
7520.0 & 8000.0 & 480.0 & 16.67 $\pm$ 0.02 \\ 
8249.0 & 8729.0 & 480.0 & 16.93 $\pm$ 0.03 \\ 
8918.0 & 9398.0 & 480.0 & 17.00 $\pm$ 0.03 \\ 
9714.0 & 10194.0 & 480.0 & 17.03 $\pm$ 0.03 \\ 
10405.0 & 10885.0 & 480.0 & 17.14 $\pm$ 0.02 \\ 
10986.0 & 11466.0 & 480.0 & 17.13 $\pm$ 0.02 \\ 
11719.0 & 12199.0 & 480.0 & 17.22 $\pm$ 0.02 \\ 
12308.0 & 12788.0 & 480.0 & 17.32 $\pm$ 0.02 \\ 
12909.0 & 13389.0 & 480.0 & 17.39 $\pm$ 0.02 \\ 
13474.0 & 13954.0 & 480.0 & 17.42 $\pm$ 0.03 \\ 
14089.0 & 14569.0 & 480.0 & 17.52 $\pm$ 0.03 \\ 
14729.0 & 15209.0 & 480.0 & 17.67 $\pm$ 0.04 \\ 
 \end{tabular}
\end{table}

\begin{table}
    \centering
	\contcaption{COATLI binned optical photometry for GRB 191016A.} 
	\label{tab:COATLI-photometry_02}
	\begin{tabular}{cccc} 
	\hline
$t_{i}$ (s) & $t_{f}$ (s) & $t_{exp}$ (s) & \emph{w} (mag) \\ 
	\hline
15419.0 & 15899.0 & 480.0 & 17.66 $\pm$ 0.03 \\ 
16133.0 & 16613.0 & 480.0 & 17.72 $\pm$ 0.03 \\ 
16730.0 & 17210.0 & 480.0 & 17.69 $\pm$ 0.03 \\ 
17372.0 & 17852.0 & 480.0 & 17.85 $\pm$ 0.04 \\ 
17972.0 & 18452.0 & 480.0 & 17.95 $\pm$ 0.04 \\ 
18555.0 & 19035.0 & 480.0 & 17.82 $\pm$ 0.04 \\ 
19330.0 & 0.0 & 480.0 & 17.97 $\pm$ 0.05 \\ 
80217.7 & 99267.7 & 19050.0 & 20.95 $\pm$ 0.07 \\ 
\hline
        \multicolumn{4}{c}{\textbf{NOTES:} t$_{i}$ and t$_{f}$ are times after \emph{Swift/BAT} trigger in seconds.} \\
    \end{tabular}
\end{table}


\begin{table}
    \vspace{1.0cm}
	\centering
	\caption{DDOTI binned optical photometry for GRB 191016A.} 
	\label{tab:DDOTI-photometry_01}
	\begin{tabular}{ccc} 
	\hline
$t_{i}$ (s)	& 	$t_{exp}$ (s)	&	 \emph{w} (mag)	\\	
\hline
765	    &	120	&	16.55$\pm$0.07	\\
1017	&	120	&	16.32$\pm$0.06	\\
1355	&	120	&	16.15$\pm$0.05	\\
2000	&	240	&	16.68$\pm$0.06	\\
2578	&	240	&	16.84$\pm$0.07	\\
3917	&	240	&	17.30$\pm$0.11	\\
5257	&	240	&	17.56$\pm$0.13	\\
5835	&	240	&	17.80$\pm$0.15	\\
7175	&	240	&	17.80$\pm$0.17	\\
8513	&	240	&	17.71$\pm$0.18	\\
9090	&	240	&	17.82$\pm$0.17	\\
10428	&	240	&	17.75$\pm$0.15	\\
11769	&	240	&	18.11$\pm$0.21	\\
12347	&	240	&	18.40$\pm$0.27	\\
13507	&	240	&	18.29$\pm$0.23	\\
14306	&	240	&	18.12$\pm$0.18	\\
15035	&	240	&	19.07$\pm$0.42	\\
16109	&	720	&	18.77$\pm$0.19	\\
17737	&	720	&	19.04$\pm$0.29	\\
19401	&	720	&	19.12$\pm$0.27	\\
\hline
        \multicolumn{3}{l}l{\textbf{NOTES:} t$_{i}$ are times after \emph{Swift/BAT} trigger in seconds.} \\
        \multicolumn{3}{l}{ t$_{exp}$ indicates the total exposure 
        time on the images included} \\
        \multicolumn{3}{l}{in the different time bins: 120(2 images), 240(4 images) and} \\
        \multicolumn{3}{l}{720(16 images), respectively.} \\
        
    \end{tabular}
\end{table}


\begin{table}
	\centering
	\caption{RATIR optical photometry for GRB 191016A.} 
	\label{tab:RATIR-photometry_01}
	\resizebox{0.45\textwidth}{!}{
	\begin{tabular}{ccccc} 
\hline
$t_{i}$ (s) & $t_{f}$ (s) & \emph{g} (mag) & \emph{r} (mag) & \emph{i} (mag) \\ 
\hline
446.4 & 526.4 & ... $\pm$ ... & 16.78 $\pm$ 0.02 & 16.48 $\pm$ 0.02 \\
837.3 & 917.3 & ... $\pm$ ... & 16.02 $\pm$ 0.01 & 15.74 $\pm$ 0.01 \\
936.4 & 1016.4 & ... $\pm$ ... & 15.86 $\pm$ 0.01	&	15.60$\pm$0.01	\\
1250.8	&	1330.8 & ...$\pm$ ... & 15.38$\pm$0.01	&	15.06$\pm$0.01	\\
1345.3	&	1425.3& ... $\pm$ ... &15.30$\pm$0.01	&	14.99$\pm$0.01	\\
1720.9	&	1800.9& ... $\pm$ ... &15.44$\pm$0.01	&	15.12$\pm$0.01	\\
1814.7	&	1894.7& ... $\pm$ ... &15.59$\pm$0.01	&	15.25$\pm$0.01	\\
2113.7	&	2193.7& ... $\pm$ ... &15.75$\pm$0.01	&	15.43$\pm$0.01	\\
2231.2	&	2311.2& ... $\pm$ ... &15.82$\pm$0.01	&	15.49$\pm$0.01	\\
2605.7	&	2685.7& ... $\pm$ ... &16.17$\pm$0.02	&	15.81$\pm$0.02	\\
2774.0	&	2854.0& ... $\pm$ ... &16.31$\pm$0.02	&	15.90$\pm$0.02	\\
3054.9	&	3134.9& ... $\pm$ ... &16.38$\pm$0.02	&	16.08$\pm$0.02	\\
3159.3	&	3239.3& ... $\pm$ ... &16.38$\pm$0.02	&	16.08$\pm$0.02	\\
3335.7	&	3415.7& ... $\pm$ ... &16.55$\pm$0.03	&	16.16$\pm$0.02	\\
3433.0	&	3513.0	& ... $\pm$ ... & ... $\pm$ ... &	16.24$\pm$0.02	\\
3545.7	&	3625.7& ... $\pm$ ... &16.63$\pm$0.03	&	16.29$\pm$0.03	\\
3663.3	&	3743.3& ... $\pm$ ... &16.62$\pm$0.03	&	16.29$\pm$0.03	\\
3757.0	&	3837.0& ... $\pm$ ... &16.72$\pm$0.03	&	16.39$\pm$0.03	\\
3855.5	&	3935.5& ... $\pm$ ... &16.72$\pm$0.04	&	16.40$\pm$0.03	\\
3948.2	&	4028.2& ... $\pm$ ... &16.68$\pm$0.03	&	...$\pm$...	\\
4054.4	&	4134.4& ... $\pm$ ... &16.89$\pm$0.05	&	16.67$\pm$0.05	\\
4170.4	&	4250.4& ... $\pm$ ... &16.85$\pm$0.04	&	16.61$\pm$0.04	\\
4265.8	&	4345.8& ... $\pm$ ... &17.00$\pm$0.06	&	16.56$\pm$0.05	\\
4370.4	&	4450.4& ... $\pm$ ... &16.95$\pm$0.06	&	16.68$\pm$0.05	\\
4470.1	&	4550.1& ... $\pm$ ... &16.93$\pm$0.05	&	16.69$\pm$0.04	\\
4577.1	&	4657.1& ... $\pm$ ... &16.92$\pm$0.04	&	16.69$\pm$0.04	\\
4676.1	&	4756.1& ... $\pm$ ... &17.04$\pm$0.06	&	17.28$\pm$0.05	\\
4780.6	&	4860.6& ... $\pm$ ... &17.65$\pm$0.01	&	16.77$\pm$0.04	\\
4969.3	&	5049.3	& ... $\pm$ ... & ... $\pm$ ... &	16.81$\pm$0.05	\\
5059.6	&	5139.6& ... $\pm$ ... &17.12$\pm$0.06	&	16.87$\pm$0.05	\\
5156.5	&	5236.5& ... $\pm$ ... &17.01$\pm$0.04	&	16.78$\pm$0.04	\\
5267.5	&	5347.5& ... $\pm$ ... &17.26$\pm$0.06	&	...$\pm$...	\\
5376.5	&	5456.5& ... $\pm$ ... &17.26$\pm$0.05	&	16.76$\pm$0.04	\\
5468.1	&	5548.1& ... $\pm$ ... &17.37$\pm$0.05	&	16.97$\pm$0.04	\\
5573.3	&	5653.3& ... $\pm$ ... &17.11$\pm$0.04	&	16.87$\pm$0.03	\\
5668.0	&	5748.0& ... $\pm$ ... &17.29$\pm$0.05	&	17.54$\pm$0.05	\\
5773.3	&	5853.3& ... $\pm$ ... &17.12$\pm$0.05	&	16.86$\pm$0.04	\\
6467.0	&	6547.0& ... $\pm$ ... &17.15$\pm$0.03	&	16.78$\pm$0.03	\\
6581.4	&	6661.4	&	17.66$\pm$0.05& ... $\pm$ ... &16.86$\pm$0.03	\\
6674.5	&	6754.5	&	17.82$\pm$0.06& ... $\pm$ ... &16.93$\pm$0.04	\\
6793.1	&	6873.1& ... $\pm$ ... &17.27$\pm$0.03	&	16.88$\pm$0.03	\\
6885.1	&	6965.1& ... $\pm$ ... &17.22$\pm$0.03	&	16.97$\pm$0.03	\\
6997.7	&	7077.7	&	17.81$\pm$0.05& ... $\pm$ ... &16.98$\pm$0.03	\\
7103.4	&	7183.4	&	18.53$\pm$0.03& ... $\pm$ ... &16.98$\pm$0.04	\\
7227.1	&	7307.1& ... $\pm$ ... &17.37$\pm$0.04	&	16.93$\pm$0.03	\\
7316.6	&	7396.6& ... $\pm$ ... &17.17$\pm$0.03	&	16.83$\pm$0.03	\\
7508.9	&	7588.9	&	17.78$\pm$0.07& ... $\pm$ ... &16.99$\pm$0.05	\\
7685.1	&	7765.1	& ... $\pm$ ... & ... $\pm$ ... &	17.38$\pm$0.06	\\
7810.1	&	7890.1& ... $\pm$ ... &17.24$\pm$0.04	&	16.92$\pm$0.03	\\
7900.9	&	7980.9& ... $\pm$ ... &17.30$\pm$0.04	&	16.98$\pm$0.03	\\
8015.6	&	8095.6	&	18.51$\pm$0.10& ... $\pm$ ... &17.12$\pm$0.04	\\
8180.6	&	8260.6	&	18.06$\pm$0.06& ... $\pm$ ... &17.04$\pm$0.04	\\
8302.5	&	8382.5& ... $\pm$ ... &17.32$\pm$0.03	&	16.97$\pm$0.03	\\
8419.8	&	8499.8& ... $\pm$ ... &17.38$\pm$0.04	&	17.04$\pm$0.03	\\
8530.1	&	8610.1	&	17.90$\pm$0.05& ... $\pm$ ... &17.07$\pm$0.04	\\
8682.0	&	8762.0& ... $\pm$ ... &17.46$\pm$0.04	&	17.17$\pm$0.04	\\
8793.0	&	8873.0& ... $\pm$ ... &17.29$\pm$0.05	&	17.05$\pm$0.05	\\
9127.5	&	9207.5& ... $\pm$ ... &17.44$\pm$0.04	&	16.89$\pm$0.03	\\
9235.5	&	9315.5	&	17.92$\pm$0.05& ... $\pm$ ... &17.11$\pm$0.03	\\
9324.7	&	9404.7	&	18.08$\pm$0.06& ... $\pm$ ... &17.17$\pm$0.04	\\
9462.1	&	9542.1& ... $\pm$ ... &17.51$\pm$0.04	&	17.13$\pm$0.04	\\
9550.5	&	9630.5& ... $\pm$ ... &17.46$\pm$0.04	&	17.21$\pm$0.03	\\
9677.1	&	9757.1	&	17.95$\pm$0.05& ... $\pm$ ... &17.22$\pm$0.04	\\
9777.1	&	9857.1	&	18.09$\pm$0.05& ... $\pm$ ... &17.19$\pm$0.04	\\
\end{tabular}}
\end{table}

\begin{table}
	\centering
	\contcaption{RATIR optical photometry for GRB 191016A.} 
	\label{tab:RATIR-photometry_02}
	\resizebox{0.45\textwidth}{!}{
	\begin{tabular}{ccccc} 
\hline
$t_{i}$ (s) & $t_{f}$ (s) & \emph{g} (mag) & \emph{r} (mag) & \emph{i} (mag) \\ 
\hline
9919.5	&	9999.5& ... $\pm$ ... &17.50$\pm$0.04	&	17.17$\pm$0.04	\\
10007.0	&	10087.0& ... $\pm$ ... &17.45$\pm$0.05	&	17.07$\pm$0.04	\\
10115.9	&	10195.9	&	18.05$\pm$0.05& ... $\pm$ ... &17.11$\pm$0.03	\\
10336.5	&	10416.5& ... $\pm$ ... &17.52$\pm$0.03	&	17.18$\pm$0.03	\\
10426.4	&	10506.4& ... $\pm$ ... &17.52$\pm$0.03	&	17.14$\pm$0.03	\\
10536.5	&	10616.5	&	18.05$\pm$0.05& ... $\pm$ ... &17.20$\pm$0.03	\\
10624.0	&	10704.0	&	18.16$\pm$0.05& ... $\pm$ ... &17.17$\pm$0.03	\\
10749.9	&	10829.9& ... $\pm$ ... &17.51$\pm$0.03	&	17.18$\pm$0.03	\\
10873.2	&	10953.2& ... $\pm$ ... &17.58$\pm$0.04	&	17.18$\pm$0.03	\\
10985.1	&	11065.1	&	18.10$\pm$0.05& ... $\pm$ ... &17.24$\pm$0.03	\\
11077.6	&	11157.6	&	18.20$\pm$0.05& ... $\pm$ ... &17.31$\pm$0.04	\\
11216.6	&	11296.6& ... $\pm$ ... &17.61$\pm$0.04	&	17.21$\pm$0.03	\\
11306.9	&	11386.9& ... $\pm$ ... &17.65$\pm$0.04	&	17.31$\pm$0.03	\\
11418.8	&	11498.8	&	18.12$\pm$0.05& ... $\pm$ ... &17.39$\pm$0.04	\\
11519.3	&	11599.3	&	18.25$\pm$0.06& ... $\pm$ ... &17.35$\pm$0.03	\\
11638.0	&	11718.0& ... $\pm$ ... &17.68$\pm$0.04	&	17.32$\pm$0.03	\\
11729.6	&	11809.6& ... $\pm$ ... &17.61$\pm$0.04	&	17.37$\pm$0.04	\\
11857.7	&	11937.7	&	18.33$\pm$0.05& ... $\pm$ ... &17.34$\pm$0.03	\\
12090.6	&	12170.6& ... $\pm$ ... &17.78$\pm$0.05	&	17.42$\pm$0.04	\\
12184.4	&	12264.4& ... $\pm$ ... &17.73$\pm$0.04	&	17.38$\pm$0.04	\\
12296.3	&	12376.3	&	18.35$\pm$0.06& ... $\pm$ ... &17.38$\pm$0.04	\\
12502.3	&	12582.3& ... $\pm$ ... &17.75$\pm$0.05	&	17.52$\pm$0.05	\\
12613.6	&	12693.6& ... $\pm$ ... &17.87$\pm$0.05	&	17.45$\pm$0.04	\\
12732.1	&	12812.1	&	18.45$\pm$0.06& ... $\pm$ ... &17.47$\pm$0.04	\\
12821.0	&	12901.0	&	18.34$\pm$0.06& ... $\pm$ ... &17.66$\pm$0.04	\\
12940.5	&	13020.5& ... $\pm$ ... &17.75$\pm$0.04	&	17.48$\pm$0.04	\\
13028.5	&	13108.5& ... $\pm$ ... &17.82$\pm$0.04	&	17.57$\pm$0.04	\\
13149.1	&	13229.1	&	18.45$\pm$0.06& ... $\pm$ ... &17.57$\pm$0.04	\\
13251.7	&	13331.7	& ... $\pm$ ... & ... $\pm$ ... &	17.85$\pm$0.06	\\
13373.1	&	13453.1& ... $\pm$ ... &17.92$\pm$0.05	&	17.60$\pm$0.04	\\
13463.0	&	13543.0& ... $\pm$ ... &17.89$\pm$0.04	&	17.60$\pm$0.04	\\
13573.6	&	13653.6	&	18.59$\pm$0.07& ... $\pm$ ... &17.69$\pm$0.04	\\
13673.7	&	13753.7	&	18.94$\pm$0.10& ... $\pm$ ... &17.61$\pm$0.04	\\
13792.1	&	13872.1& ... $\pm$ ... &17.96$\pm$0.05	&	17.63$\pm$0.04	\\
13888.1	&	13968.1& ... $\pm$ ... &17.98$\pm$0.05	&	17.62$\pm$0.04	\\
13997.0	&	14077.0	&	18.77$\pm$0.08& ... $\pm$ ... &17.77$\pm$0.04	\\
14084.8	&	14164.8	&	18.84$\pm$0.08& ... $\pm$ ... &17.68$\pm$0.04	\\
14203.2	&	14283.2& ... $\pm$ ... &17.94$\pm$0.04	&	17.63$\pm$0.04	\\
14734.2	&	14814.2& ... $\pm$ ... &17.95$\pm$0.04	&	17.68$\pm$0.04	\\
14847.3	&	14927.3	&	18.68$\pm$0.07& ... $\pm$ ... &17.76$\pm$0.04	\\
14936.5	&	15016.5	&	18.61$\pm$0.06& ... $\pm$ ... &17.80$\pm$0.04	\\
15057.2	&	15137.2& ... $\pm$ ... &18.13$\pm$0.05	&	17.73$\pm$0.04	\\
15165.8	&	15245.8& ... $\pm$ ... &17.93$\pm$0.04	&	17.76$\pm$0.04	\\
15276.6	&	15356.6	&	18.72$\pm$0.07& ... $\pm$ ... &17.69$\pm$0.04	\\
15376.8	&	15456.8	&	18.97$\pm$0.09& ... $\pm$ ... &17.85$\pm$0.05	\\
15516.6	&	15596.6& ... $\pm$ ... &18.08$\pm$0.05	&	17.78$\pm$0.04	\\
15608.8	&	15688.8& ... $\pm$ ... &18.21$\pm$0.05	&	17.87$\pm$0.04	\\
15719.3	&	15799.3	&	18.76$\pm$0.07& ... $\pm$ ... &17.87$\pm$0.05	\\
15964.2	&	16044.2& ... $\pm$ ... &18.03$\pm$0.05	&	17.76$\pm$0.04	\\
16053.2	&	16133.2& ... $\pm$ ... &18.11$\pm$0.05	&	17.80$\pm$0.04	\\
16378.3	&	16458.3& ... $\pm$ ... &18.19$\pm$0.05	&	17.91$\pm$0.05	\\
16494.9	&	16574.9& ... $\pm$ ... &18.23$\pm$0.06	&	17.85$\pm$0.05	\\
16605.3	&	16685.3	&	18.87$\pm$0.08& ... $\pm$ ... &17.84$\pm$0.04	\\
16695.3	&	16775.3	&	18.89$\pm$0.09& ... $\pm$ ... &17.92$\pm$0.05	\\
16815.5	&	16895.5& ... $\pm$ ... &18.39$\pm$0.06	&	17.95$\pm$0.05	\\
16908.7	&	16988.7& ... $\pm$ ... &18.24$\pm$0.06	&	17.98$\pm$0.05	\\
17027.9	&	17107.9	&	18.78$\pm$0.08& ... $\pm$ ... &17.96$\pm$0.05	\\
17128.8	&	17208.8	&	18.77$\pm$0.07& ... $\pm$ ... &17.95$\pm$0.05	\\
17248.6	&	17328.6& ... $\pm$ ... &18.19$\pm$0.05	&	17.89$\pm$0.04	\\
17336.4	&	17416.4& ... $\pm$ ... &18.37$\pm$0.06	&	18.05$\pm$0.05	\\
81362.5	&	86962.5& ... $\pm$ ... &21.18$\pm$0.15	&	20.99$\pm$0.08	\\
81471.4	&	92991.4	&	22.19$\pm$0.21& ... $\pm$ ... &...$\pm$...	\\
167974.1	&	179254.1& ... $\pm$ ... &22.09$\pm$0.00	&	22.56$\pm$0.24	\\
\hline
        \multicolumn{5}{c}{\textbf{NOTES:} t$_{i}$ and t$_{f}$ are times after \emph{Swift/BAT} trigger in seconds.} \\
    \end{tabular}}
\end{table}

\begin{table}
	\caption{RATIR NIR photometry for GRB 191016A.} 
	\label{tab:RATIR-NIR_01}
	\resizebox{0.5\textwidth}{!}{
	\begin{tabular}{cccccc} 
\hline
$t_{i}$ (s) & $t_{f}$ (s) & \emph{Z} (mag) & \emph{Y} (mag) & \emph{J} (mag) & \emph{H} (mag) \\ 
\hline
465.6	&	535.6	&	16.31	$\pm$ 	0.03	&	 ... 	$\pm$ 	 ... 	&	16.07	$\pm$ 	0.03	&	 ...	$\pm$ 	 ...	\\
582.6	&	651.6	&	 ... 	$\pm$ 	 ... 	&	15.92	$\pm$ 	0.02	&	 ... 	$\pm$ 	 ... 	&	15.52	$\pm$ 	0.03	\\
671.6	&	738.6	&	 ... 	$\pm$ 	 ... 	&	15.72	$\pm$ 	0.02	&	 ... 	$\pm$ 	 ... 	&	 ...	$\pm$ 	 ...	\\
750.6	&	817.6	&	 ... 	$\pm$ 	 ... 	&	 ... 	$\pm$ 	 ... 	&	 ... 	$\pm$ 	 ... 	&	15.15	$\pm$ 	0.02	\\
852.6	&	920.6	&	15.47	$\pm$ 	0.02	&	 ... 	$\pm$ 	 ... 	&	15.17	$\pm$ 	0.02	&	 ...	$\pm$ 	 ...	\\
949.6	&	1021.6	&	15.37	$\pm$ 	0.02	&	 ... 	$\pm$ 	 ... 	&	15.03	$\pm$ 	0.01	&	 ...	$\pm$ 	 ...	\\
1048.6	&	1120.6	&	 ... 	$\pm$ 	 ... 	&	15.00	$\pm$ 	0.01	&	 ... 	$\pm$ 	 ... 	&	14.52	$\pm$ 	0.01	\\
1161.6	&	1231.6	&	 ... 	$\pm$ 	 ... 	&	14.70	$\pm$ 	0.01	&	 ... 	$\pm$ 	 ... 	&	14.24	$\pm$ 	0.01	\\
1266.6	&	1336.6	&	14.79	$\pm$ 	0.01	&	 ... 	$\pm$ 	 ... 	&	14.42	$\pm$ 	0.01	&	 ...	$\pm$ 	 ...	\\
1357.6	&	1424.6	&	 ... 	$\pm$ 	 ... 	&	 ... 	$\pm$ 	 ... 	&	14.33	$\pm$ 	0.01	&	 ...	$\pm$ 	 ...	\\
1434.6	&	1501.6	&	14.70	$\pm$ 	0.01	&	 ... 	$\pm$ 	 ... 	&	 ... 	$\pm$ 	 ... 	&	 ...	$\pm$ 	 ...	\\
1527.6	&	1597.6	&	 ... 	$\pm$ 	 ... 	&	14.52	$\pm$ 	0.01	&	 ... 	$\pm$ 	 ... 	&	14.02	$\pm$ 	0.01	\\
1627.6	&	1700.6	&	 ... 	$\pm$ 	 ... 	&	14.58	$\pm$ 	0.01	&	 ... 	$\pm$ 	 ... 	&	14.07	$\pm$ 	0.01	\\
1735.6	&	1802.6	&	 ... 	$\pm$ 	 ... 	&	 ... 	$\pm$ 		&	14.43	$\pm$ 	0.01	&	 ...	$\pm$ 	 ...	\\
1933.6	&	2004.6	&	 ... 	$\pm$ 	 ... 	&	14.83	$\pm$ 	0.01	&	 ... 	$\pm$ 	 ... 	&	14.33	$\pm$ 	0.01	\\
2026.6	&	2096.6	&	 ... 	$\pm$ 	 ... 	&	14.92	$\pm$ 	0.01	&	 ... 	$\pm$ 	 ... 	&	14.42	$\pm$ 	0.01	\\
2128.6	&	2195.6	&	 ... 	$\pm$ 	 ... 	&	 ... 	$\pm$ 		&	14.73	$\pm$ 	0.01	&	 ...	$\pm$ 	 ...	\\
2244.6	&	2314.6	&	15.24	$\pm$ 	0.01	&	 ... 	$\pm$ 		&	14.84	$\pm$ 	0.01	&	 ...	$\pm$ 	 ...	\\
2341.6	&	2408.6	&	 ... 	$\pm$ 	 ... 	&	15.16	$\pm$ 	0.01	&	 ... 	$\pm$ 	 ... 	&	 ...	$\pm$ 	 ...	\\
2420.6	&	2487.6	&	 ... 	$\pm$ 	 ... 	&	 ... 	$\pm$ 		&	 ... 	$\pm$ 	 ... 	&	14.68	$\pm$ 	0.01	\\
2512.6	&	2580.6	&	 ... 	$\pm$ 	 ... 	&	15.27	$\pm$ 	0.01	&	 ... 	$\pm$ 	 ... 	&	14.73	$\pm$ 	0.01	\\
2696.6	&	2763.6	&	 ... 	$\pm$ 	 ... 	&	 ... 	$\pm$ 		&	15.41	$\pm$ 	0.02	&	 ...	$\pm$ 	 ...	\\
2788.6	&	2855.6	&	 ... 	$\pm$ 	 ... 	&	 ... 	$\pm$ 		&	15.37	$\pm$ 	0.02	&	 ...	$\pm$ 	 ...	\\
2884.6	&	2951.6	&	 ... 	$\pm$ 	 ... 	&	 ... 	$\pm$ 		&	 ... 	$\pm$ 	 ... 	&	14.96	$\pm$ 	0.01	\\
2964.6	&	3031.6	&	 ... 	$\pm$ 	 ... 	&	15.53	$\pm$ 	0.02	&	 ... 	$\pm$ 	 ... 	&	 ...	$\pm$ 	 ...	\\
3070.6	&	3140.6	&	 ... 	$\pm$ 	 ... 	&	15.57	$\pm$ 	0.02	&	 ... 	$\pm$ 	 ... 	&	15.10	$\pm$ 	0.02	\\
3173.6	&	3240.6	&	 ... 	$\pm$ 	 ... 	&	 ... 	$\pm$ 		&	15.42	$\pm$ 	0.02	&	 ...	$\pm$ 	 ...	\\
3253.6	&	3320.6	&	15.91	$\pm$ 	0.02	&	 ... 	$\pm$ 		&	 ... 	$\pm$ 	 ... 	&	 ...	$\pm$ 	 ...	\\
3350.6	&	3420.6	&	16.01	$\pm$ 	0.03	&	 ... 	$\pm$ 		&	15.52	$\pm$ 	0.02	&	 ...	$\pm$ 	 ...	\\
3447.6	&	3519.6	&	 ... 	$\pm$ 	 ... 	&	15.75	$\pm$ 	0.02	&	 ... 	$\pm$ 	 ... 	&	15.25	$\pm$ 	0.02	\\
3560.6	&	3630.6	&	 ... 	$\pm$ 	 ... 	&	15.74	$\pm$ 	0.02	&	 ... 	$\pm$ 	 ... 	&	15.26	$\pm$ 	0.02	\\
3677.6	&	3748.6	&	16.06	$\pm$ 	0.03	&	 ... 	$\pm$ 		&	15.63	$\pm$ 	0.02	&	 ...	$\pm$ 	 ...	\\
3771.6	&	3843.6	&	16.09	$\pm$ 	0.03	&	 ... 	$\pm$ 		&	15.71	$\pm$ 	0.02	&	 ...	$\pm$ 	 ...	\\
3869.6	&	3940.6	&	 ... 	$\pm$ 	 ... 	&	15.90	$\pm$ 	0.02	&	 ... 	$\pm$ 	 ... 	&	15.44	$\pm$ 	0.02	\\
3962.6	&	4031.6	&	 ... 	$\pm$ 	 ... 	&	15.89	$\pm$ 	0.02	&	 ... 	$\pm$ 	 ... 	&	15.48	$\pm$ 	0.02	\\
4069.6	&	4137.6	&	16.29	$\pm$ 	0.04	&	 ... 	$\pm$ 		&	15.79	$\pm$ 	0.03	&	 ...	$\pm$ 	 ...	\\
4187.6	&	4254.6	&	16.34	$\pm$ 	0.04	&	 ... 	$\pm$ 		&	15.86	$\pm$ 	0.03	&	 ...	$\pm$ 	 ...	\\
4279.6	&	4349.6	&	 ... 	$\pm$ 	 ... 	&	16.04	$\pm$ 	0.03	&	 ... 	$\pm$ 	 ... 	&	15.53	$\pm$ 	0.03	\\
4384.6	&	4451.6	&	 ... 	$\pm$ 	 ... 	&	16.12	$\pm$ 	0.04	&	 ... 	$\pm$ 	 ... 	&	15.69	$\pm$ 	0.03	\\
4483.6	&	4555.6	&	16.35	$\pm$ 	0.04	&	 ... 	$\pm$ 		&	15.97	$\pm$ 	0.03	&	 ...	$\pm$ 	 ...	\\
4591.6	&	4662.6	&	16.44	$\pm$ 	0.05	&	 ... 	$\pm$ 		&	15.94	$\pm$ 	0.03	&	 ...	$\pm$ 	 ...	\\
4690.6	&	4761.6	&	 ... 	$\pm$ 	 ... 	&	16.20	$\pm$ 	0.03	&	 ... 	$\pm$ 	 ... 	&	15.68	$\pm$ 	0.03	\\
4796.6	&	4863.6	&	 ... 	$\pm$ 	 ... 	&	16.20	$\pm$ 	0.03	&	 ... 	$\pm$ 	 ... 	&	 ...	$\pm$ 	 ...	\\
4874.6	&	4941.6	&	 ... 	$\pm$ 	 ... 	&	 ... 	$\pm$ 		&	 ... 	$\pm$ 	 ... 	&	15.73	$\pm$ 	0.03	\\
4983.6	&	5051.6	&	16.54	$\pm$ 	0.04	&	 ... 	$\pm$ 		&	16.00	$\pm$ 	0.03	&	 ...	$\pm$ 	 ...	\\
5076.6	&	5143.6	&	 ... 	$\pm$ 	 ... 	&	 ... 	$\pm$ 		&	15.97	$\pm$ 	0.03	&	 ...	$\pm$ 	 ...	\\
5171.6	&	5242.6	&	 ... 	$\pm$ 	 ... 	&	16.32	$\pm$ 	0.03	&	 ... 	$\pm$ 	 ... 	&	15.74	$\pm$ 	0.03	\\
5283.6	&	5355.6	&	 ... 	$\pm$ 	 ... 	&	16.32	$\pm$ 	0.03	&	 ... 	$\pm$ 	 ... 	&	15.78	$\pm$ 	0.03	\\
5481.6	&	5553.6	&	16.55	$\pm$ 	0.04	&	 ... 	$\pm$ 		&	16.09	$\pm$ 	0.03	&	 ...	$\pm$ 	 ...	\\
5587.6	&	5659.6	&	 ... 	$\pm$ 	 ... 	&	16.41	$\pm$ 	0.03	&	 ... 	$\pm$ 	 ... 	&	15.92	$\pm$ 	0.03	\\
5681.6	&	5751.6	&	 ... 	$\pm$ 	 ... 	&	16.40	$\pm$ 	0.03	&	 ... 	$\pm$ 	 ... 	&	15.94	$\pm$ 	0.03	\\
5789.6	&	5856.6	&	16.60	$\pm$ 	0.05	&	 ... 	$\pm$ 		&	16.21	$\pm$ 	0.03	&	 ...	$\pm$ 	 ...	\\
6486.6	&	6554.6	&	16.60	$\pm$ 	0.04	&	 ... 	$\pm$ 		&	16.13	$\pm$ 	0.03	&	 ...	$\pm$ 	 ...	\\
6598.6	&	6667.6	&	 ... 	$\pm$ 	 ... 	&	16.35	$\pm$ 	0.03	&	 ... 	$\pm$ 	 ... 	&	15.91	$\pm$ 	0.02	\\
6689.6	&	6756.6	&	 ... 	$\pm$ 	 ... 	&	16.51	$\pm$ 	0.03	&	 ... 	$\pm$ 	 ... 	&	15.91	$\pm$ 	0.02	\\
6808.6	&	6878.6	&	16.65	$\pm$ 	0.04	&	 ... 	$\pm$ 		&	16.29	$\pm$ 	0.03	&	 ...	$\pm$ 	 ...	\\
6900.6	&	6969.6	&	16.74	$\pm$ 	0.04	&	 ... 	$\pm$ 		&	16.32	$\pm$ 	0.03	&	 ...	$\pm$ 	 ...	\\
7013.6	&	7082.6	&	 ... 	$\pm$ 	 ... 	&	16.53	$\pm$ 	0.03	&	 ... 	$\pm$ 	 ... 	&	16.05	$\pm$ 	0.03	\\
7117.6	&	7186.6	&	 ... 	$\pm$ 	 ... 	&	16.45	$\pm$ 	0.03	&	 ... 	$\pm$ 	 ... 	&	16.04	$\pm$ 	0.03	\\
7240.6	&	7310.6	&	16.66	$\pm$ 	0.04	&	 ... 	$\pm$ 		&	16.26	$\pm$ 	0.03	&	 ...	$\pm$ 	 ...	\\
7328.6	&	7395.6	&	16.67	$\pm$ 	0.04	&	 ... 	$\pm$ 		&	 ... 	$\pm$ 	 ... 	&	 ...	$\pm$ 	 ...	\\
7406.6	&	7473.6	&	 ... 	$\pm$ 	 ... 	&	 ... 	$\pm$ 		&	16.22	$\pm$ 	0.03	&	 ...	$\pm$ 	 ...	\\
7522.6	&	7589.6	&	 ... 	$\pm$ 	 ... 	&	16.38	$\pm$ 	0.03	&	 ... 	$\pm$ 	 ... 	&	 ...	$\pm$ 	 ...	\\
7601.6	&	7668.6	&	 ... 	$\pm$ 	 ... 	&	 ... 	$\pm$ 		&	 ... 	$\pm$ 	 ... 	&	15.97	$\pm$ 	0.03	\\
7701.6	&	7772.6	&	 ... 	$\pm$ 	 ... 	&	16.47	$\pm$ 	0.03	&	 ... 	$\pm$ 	 ... 	&	15.99	$\pm$ 	0.03	\\
7823.6	&	7893.6	&	16.70	$\pm$ 	0.05	&	 ... 	$\pm$ 		&	16.23	$\pm$ 	0.03	&	 ...	$\pm$ 	 ...	\\
7914.7	&	7982.7	&	16.67	$\pm$ 	0.05	&	 ... 	$\pm$ 		&	16.36	$\pm$ 	0.04	&	 ...	$\pm$ 	 ...	\\
8107.7	&	8174.7	&	 ... 	$\pm$ 	 ... 	&	16.54	$\pm$ 	0.04	&	 ... 	$\pm$ 	 ... 	&	 ...	$\pm$ 	 ...	\\
8195.7	&	8262.7	&	 ... 	$\pm$ 	 ... 	&	16.57	$\pm$ 	0.04	&	 ... 	$\pm$ 	 ... 	&	 ...	$\pm$ 	 ...	\\
8319.7	&	8386.7	&	16.81	$\pm$ 	0.05	&	 ... 	$\pm$ 		&	 ... 	$\pm$ 	 ... 	&	 ...	$\pm$ 	 ...	\\
8435.7	&	8502.7	&	16.90	$\pm$ 	0.06	&	 ... 	$\pm$ 		&	 ... 	$\pm$ 	 ... 	&	 ...	$\pm$ 	 ...	\\
\hline
        \multicolumn{6}{c}{\textbf{NOTES:} t$_{i}$ and t$_{f}$ are times after \emph{Swift/BAT} trigger in seconds.} \\
\end{tabular}}
\end{table}


\bsp	
\label{lastpage}
\end{document}